\documentclass[12pt,hang,flushmargin]{article}
\usepackage[
    margin=0.9in
    ]{geometry}
\usepackage[utf8]{inputenc}
\usepackage[english]{babel}
\usepackage{amsmath,amsfonts,amssymb}
\usepackage{caption,adjustbox}
\usepackage{subcaption}
\usepackage[toc]{appendix}
\usepackage{multirow}
\usepackage{fontawesome}
\usepackage{comment}
\usepackage[outdir=./]{epstopdf}
\usepackage{graphicx}
\usepackage{footmisc}
\usepackage{footnotebackref}
\usepackage{hyperref}
\usepackage{bbding}
\usepackage{pifont}
\usepackage{wasysym}
\hypersetup{
    colorlinks,
    citecolor=black,
    filecolor=black,
    linkcolor=black,
    urlcolor=black
}
\usepackage{wrapfig}
\usepackage[usenames,dvipsnames]{xcolor}
\usepackage[arrowdel]{physics}
\usepackage{tensor}
\usepackage{cancel}
\usepackage{tikz}
\usetikzlibrary{patterns}
\usepackage{cleveref}
\usepackage[affil-it]{authblk}
\usepackage{listings} 
\usepackage[style=numeric-comp,sorting=none,maxbibnames=9]{biblatex}
\newcommand{\gnewton}{G_\mathrm{N}}

\usepackage{subfiles}
\definecolor{rossoCP3}{cmyk}{0,.88,.77,.40}

\newcommand{\geo}{\mathfrak{v}}

\newcommand{\zh}{z_{H}}
\newcommand{\za}{z_A}

\title{
\begin{flushright}{\vspace{-2.5cm}\small LYCEN 2023-01\\}\end{flushright}
\vspace{2.3cm}
\LARGE\color{rossoCP3}{\textbf{
Positivity Conditions \\ [.1cm] for \\[.1cm] Generalised Schwarzschild Space-Times\\[0.4cm]
}}}
 
\author{\normalsize  Alessandra \textsc{D'Alise}$^{\star\color{rossoCP3}{\heartsuit}}$, Giuseppe \textsc{Fabiano}$^{\sharp\color{rossoCP3}{\heartsuit}}
$, Domenico \textsc{Frattulillo}$^{\wp\color{rossoCP3}{\heartsuit}}
$, Stefan \textsc{Hohenegger}$^{\circledS\blacktriangle}
$, Davide \textsc{Iacobacci}$^{\dagger\color{rossoCP3}{\heartsuit}}
$, Franco \textsc{Pezzella}$^{\circlearrowright\color{rossoCP3}{\heartsuit}}$, Francesco \textsc{Sannino}$^{\ddagger \color{rossoCP3}{\diamondsuit}}$$^{\clubsuit\color{rossoCP3}{\heartsuit}}$}

\date{}

\begin{document}

\maketitle
\vskip -2cm
{\let\thefootnote\relax\footnote{$^\star$ Electronic address: \textcolor{rossoCP3}{\href{mailto:alessandra.dalise@unina.it}{\textcolor{rossoCP3}{alessandra.dalise@unina.it}}}\\$^\sharp$ Electronic address: \textcolor{rossoCP3}{\href{mailto:giuseppe.fabiano@unina.it}{\textcolor{rossoCP3}{giuseppe.fabiano@unina.it}}}\\$^\wp$ Electronic address: \textcolor{rossoCP3}{\href{mailto:domenico.frattulillo@unina.it}{\textcolor{rossoCP3}{domenico.frattulillo@unina.it}}}\\$^\circledS$ Electronic address: \textcolor{rossoCP3}{\href{mailto:s.hohenegger@ipnl.in2p3.fr}{\textcolor{rossoCP3}{s.hohenegger@ipnl.in2p3.fr}}}\\$^\dagger$ Electronic address: \textcolor{rossoCP3}{\href{mailto:davide.iacobacci@unina.it}{\textcolor{rossoCP3}{davide.iacobacci@unina.it}}}\\$^\circlearrowright$ Electronic address: \textcolor{rossoCP3}{\href{mailto:pezzella@na.infn.it}{\textcolor{rossoCP3}{pezzella@na.infn.it}}}\\$^\ddagger$ Electronic address: \textcolor{rossoCP3}{\href{mailto:sannino@cp3.sdu.dk}{\textcolor{rossoCP3}{sannino@cp3.sdu.dk}}}}}\vspace{0.5mm} 
\begin{center}
\footnotesize{$\color{rossoCP3}{^\heartsuit}$ Dipartimento di Fisica ``E. Pancini", Università di Napoli Federico II - INFN sezione di Napoli, Complesso Universitario di Monte S. Angelo Edificio 6, via Cintia, 80126 Napoli, Italy}\\
\footnotesize{$^\blacktriangle$ Univ Lyon, Univ Claude Bernard Lyon 1, CNRS/IN2P3, IP2I Lyon, UMR 5822, F-69622, Villeurbanne, France}\\
\footnotesize{$^\clubsuit$ Scuola Superiore Meridionale, Largo S. Marcellino, 10, 80138 Napoli NA, Italy}\\
\footnotesize{$\color{rossoCP3}{^\diamondsuit}$ Quantum Field Theory Center (QTC) and D-IAS, Univ. of Southern Denmark, Campusvej 55, 5230 Odense M, Denmark}
\end{center}

\begin{abstract}
{We analyse the impact of positivity conditions on static spherically symmetric deformations of the Schwarzschild space-time. The metric is taken to satisfy, at least asymptotically, the Einstein equation in the presence of a non-trivial stress-energy tensor, on which we impose various physicality conditions. We systematically study and compare the impact of these conditions on the space-time deformations. The universal nature of our findings applies to both classical and quantum metric deformations with and without event horizons. We further discuss minimal realisations of the asymptotic stress energy tensor in terms of physical fields. Finally, we illustrate our results by discussing concrete models of quantum black holes.}
 
\end{abstract}
\newpage
\tableofcontents
\section{Introduction}

\noindent
{{The Einstein equation $G_{\mu\nu}=8\pi\, T_{\mu\nu}$ describes classical gravity coupled to a matter system. The former is represented by the Einstein tensor $G_{\mu\nu}$ that encodes geometric quantities of the space-time, which in this paper we shall consider to be 4-dimensional with Minkowski signature. The latter is represented by the energy momentum tensor ${T^\mu}_\nu$, which can be realised through various different (classical) fields. For each such configuration, the Einstein equation allows to compute the associated space-time metric. While, however, a general local classification of possible such tensors exists \cite{Hawking:1973uf}, based on the nature of its eigenvectors (see Appendix~\ref{App:EllisHawking} for a review), the properties of this metric depend strongly on more specific details of ${T^\mu}_\nu$. This makes general statements about large scale structures in the universe difficult, which in particular concerns singular space-times such as black holes, naked singularities or worm holes. Rather than studying separately solutions for explicit (classes of) energy momentum tensors, various \emph{energy conditions} \cite{Hawking:1973uf,Visser:1996iw,Ford:1994bj,Martin-Moruno:2013wfa,Curiel:2014zba,Kontou:2020bta} have been discussed that characterise physically reasonable matter systems. Generally, these conditions state that physical observers cannot measure negative energy densities and/or space-like energy flows, either point-wise or averaged over specific regions in space-time (we provide a more detailed overview in Section~\ref{Sect:ConditionsT}). Indeed, through the celebrate Raychaudhuri equation~\cite{Raychaudhuri:1953yv} (see Appendix~\ref{App:Raychaudhuri} for a review), these conditions can be translated into geometric properties, namely the behaviour of time-like and null congruences (\emph{i.e.} collections of geodesics). Analysing these equations has lead to powerful statements about the (non-)existence of singularities \cite{Penrose:1964wq,Penrose:1969pc,osti_4155937,Hawking:1970zqf,Wald:1984rg}. 

The above mentioned energy conditions are hard to derive from first principles (see \emph{e.g.} \cite{Trautman1965-TRAFAC,ATrautman_1966,PhysRevD.86.083515}) and are mostly motivated by studying (classical) examples. Many of them, however, are also known to be violated in certain situations, notably in cases where quantum effects are taken into account \cite{Epstein:1965zza,Fewster:2002ne,Visser:1996iv}. This has recently lead to a critical re-evaluation of these conditions (see \emph{e.g.} \cite{Barcelo:2002bv}): since even in the quantum case the violation of the positivity of the energy density or the space-like character of the energy flux remain in general bounded, several \emph{quantum energy inequalities} have been proposed (see Section~\ref{Sect:ConditionsT} for an overview). Other proposals for quantum conditions for the energy momentum tensor take into account the Bousso-bound \cite{Bousso:2015mna,Bousso:2015wca} and are related to the (non-)conservation of information. The classification of space-time singularities (or the demonstration of their absence) in view of this more general conditions is a challenging task. Moreover, in the absence of a complete theory of quantum gravity, the generalisation of the Einstein equation is not known.

In this paper, rather than computing the space-time metric associated to a given matter system by solving the Einstein equation (or a generalisation thereof in the case of quantum gravity), we shall start from a given static and spherically symmetric metric that asymptotically approaches the one describing the classical Schwarzschild space-time \cite{Schwarzschild}. We shall then assume that for sufficiently large distances from the rotational center, we can associate the Einstein tensor of this space-time with an energy momentum tensor made from (classical) field configurations. Studying the above mentioned energy conditions (and their quantum modifications) implies non-trivial conditions for modifications of the Schwarzschild metric. In this way, we establish guidelines for possible (quantum-) deformations of spherically symmetric and static space-time metrics.

Indeed, in recent years, various different examples of deformations of the Schwarzschild metric have been investigated \cite{dymnikova1992vacuum,Bonanno:2000ep,Bjerrum-Bohr:2002fji,Gonzalez:2015upa,Nicolini:2019irw,Ruiz:2021qfp,Hayward_2006,Platania:2019kyx,Knorr:2022kqp,Binetti:2022xdi,Eichhorn:2022bgu,bardeen1968proceedings}, stemming either from modified theories of gravity or as proposed quantum generalisations. In the recent paper \cite{Binetti:2022xdi} 
some of the current authors have proposed a general framework to describe quantum corrections to static, spherically symmetric black hole solutions. Based on the renormalisation group approach \cite{ChenGoldenfeldOono1995,,PhysRevE.49.4502,Barenblatt}, corrections to the Schwarzschild metric are encoded in functions of a physical distance $d$ (in the simplest form as an asymptotic series in inverse powers of $d$). Modifying the metric with a function depending on a physical quantity, guarantees invariance of the space-time under coordinate reparametrisations (see \cite{Held:2021vwd,} for proposals to use other physical quantities to this end). In this way, quantum-corrected space-times can be described in a model independent fashion, while still allowing to compute physically interesting quantities, such as the  temperature or entropy. Indeed, the details of the underlying model of quantum gravity are encoded in the above mentioned functions of $d$, for example the coefficients of an asymptotic expansion. Moreover, at least formally, we can associate with these space-time metrics energy momentum tensors in the sense of the Einstein equation. Assuming that this ${T^\mu}_\nu$, at least for sufficiently large distances, resembles a physically reasonable (classical) system it allows us to imply the above mentioned energy conditions. As we shall discover in this paper, this poses non-trivial conditions on the deformation functions and concretely also restrictions on the asymptotic expansion coefficients. 

Our discussion, however, is not limited to black hole space-times. 
In fact, our approach can be applied to general spherically symmetric and static geometries, that asymptotically approach the Schwarzschild metric. Since our conditions are imposed in the asymptotic regime, where we assume that the geometry can be well approximated by classical General Relativity, our approach works equally for geometries that feature event horizons and/or singularities at the origin. Furthermore, our approach can be applied to any type of deformations of the metric, either due to quantum effects or any other modifications of classical General Relativity (GR).

This paper is organised as follows: In Section~\ref{fluid hypo}, we introduce the general form of the static and spherically symmetric metric, notably its classical approximation in the asymptotic regime. This allows us to re-write the metric deformation into a stress energy tensor that appears on the right-hand side of the Einstein equation. We then review different physicality conditions that can be imposed on this energy momentum tensors: while we mostly focus on the classical case, we also mention possible quantum extensions. We further review minimal realisations of such energy momentum tensors in terms of (classical) field configurations. In Section~\ref{Sect:EnergyConditionsGen} we study the impact of the energy conditions on the metric deformations. Keeping the latter generic, we find non-trivial conditions for the deformation functions and their derivatives. In Section~\ref{Sect:AsymptoticExpansion} we specify the metric functions to asymptotic series expansions in an inverse distance function, following the discussion of \cite{Binetti:2022xdi}. In this way, we can characterise the energy conditions through allowed regions in the parameter space of the expansion coefficients. In Section~\ref{Sect:Examples} we discuss as further concrete examples the Dymnikova space-time \cite{dymnikova1992vacuum} and the Bonanno-Reuter black hole \cite{Bonanno:2000ep}. Finally, Section~\ref{Sect:Conclusions} contains our conclusions. Furthermore, technical details and reviews of relevant details have been relegate to two appendices: Appendix~\ref{App:SpaceTime} contains information about properties of spherically symmetric metrics, the classification of energy moment tensors and the Raychaudhuri equation. Appendix~\ref{App:SchemeDependence} contains a discussion about the equivalence of different choices of the physical distance.

}}

\section{Spherical and Static Space-time}\label{fluid hypo}

We consider a class of static and spherically symmetric metrics \cite{Schwarzschild} of the form
\begin{equation}\label{metrica adimensionale}
    \dd {\sigma}^2=-h(z) \dd \tau^2+\frac{\dd z^2}{f(z)}+z^2 \dd\theta^2+z^2\sin^2\theta\dd\varphi^2 = g_{\mu\nu}dx^\mu dx^\nu\ , 
\end{equation}
where the infinitesimal distance  $d\sigma$ and coordinates $\tau$,$z$ are written in units of the Planck length $\ell_p = 1/M_p$ and $M_p$ is the Planck mass. Furthermore, the dimensionless metric is $g_{\mu\nu} = {\rm diag}\{-h,1/f,z^2, z^2 \sin^2\theta \}$, which we assume to approach the Schwarzschild one \cite{Schwarzschild} for $z\rightarrow \infty$ namely
\begin{equation}
    \lim_{z\rightarrow \infty} f(z)=\lim_{z\rightarrow \infty} h(z)=1-\frac{2\chi}{z}
\end{equation}
where $\chi=\frac{M}{M_p}$ is {a dimensionless parameter (which represents the mass of the central body)}. This implies that at asymptotic large distances our metric is solution of the Einstein gravity in the presence of a static and spherical symmetric source. Furthermore for the metric to describe a black hole we further require
\begin{equation}
    \lim_{z\rightarrow z_H} f(z)=\lim_{z\rightarrow z_H} h(z)= 0 \ ,
\end{equation}
with $z_H$ the location of the (outer) event horizon, such that
\begin{align}
&f(z)>0\,,&&\text{and}&&h(z)>0\,,&&\forall z>z_H\,.\label{PosMetricFunctions}
\end{align}
More information about this geometry has been compiled in Appendix~\ref{App:Geometry}. The deformation from the classical Schwarzschild metric can arise from  either classical and/or quantum modification of Einstein gravity. We will consider explicit realisations in the following sections.



\subsection{Energy Momentum Tensor}\label{Sect:EnergyMomentum}
The functions $h$ and $f$ in (\ref{metrica adimensionale}) characterise the deformation of the space-time relative to the Schwarzschild metric and arise as solutions of some model of modified (quantum) gravity. While the exact form of this solution  depends on the model, we shall explore some of its properties from more general considerations, namely we shall examine the impact of various different \emph{energy conditions} from the literature (for more details see Section~\ref{Sect:ConditionsT}). Such conditions can be formulated purely geometrically (\emph{i.e.} in terms of the space-time metric (\ref{metrica adimensionale})) and we shall explore their impact on the functions $h$ and $f$. In most (but not all) cases, however, the physical interpretation and intuition of these energy conditions stems from the Einstein equation (see \cite{Curiel:2014zba} for a discussion of this point):
\begin{align}
&G_{\mu\nu}=8\pi\,  T_{\mu\nu} \ ,&&\text{with} &&G_{\mu \nu} = R_{\mu\nu} - R\, \frac{g_{\mu\nu}}{2}\,.\label{EQT}
\end{align}
Here $G_{\mu\nu}$ is the Einstein tensor, $R_{\mu\nu}$ and $R$ the dimensionless Ricci tensor and scalar and $T_{\mu\nu}$ a suitable energy momentum tensor. We shall assume that the metric (\ref{metrica adimensionale}), at least asymptotically (\emph{i.e.} far away from the origin\footnote{{In the following Section, we shall introduce a radius $\za>0$ beyond which we assume this condition to be satisfied.}}) can approximately be described by (\ref{EQT}) with an energy momentum tensor $T_{\mu\nu}$ that satisfies 'reasonable' conditions, which we shall outline in more detail in the following Subsection. Before, however, a few more comments are in order concerning eq.~(\ref{EQT}): 
\begin{itemize}
\item  The Einstein tensor  $G_{\mu \nu}$ is a purely geometric quantity and directly determined from the metric (\ref{metrica adimensionale}). In our approach below, eq.~(\ref{EQT}) associates to each pair of metric functions $f$ and $h$ an energy momentum tensor (see eq.~(\ref{emtensor}) below and eq.~(\ref{FormT}) in the appendix for the concrete form). In the case of the Schwarzschild geometry, \emph{i.e.} the metric in eq.~(\ref{metrica adimensionale}) with the functions
\begin{align}
&f(z)=f_S(z)=1-\frac{2\chi}{z}\,,&&\text{and} &&h(z)=h_S(z)=1-\frac{2\chi}{z}\,,\label{SchwarzschildMetric}
\end{align}
we have $T_{\mu\nu}=0$ {for $z>0$}.

\item Our approach is general in the sense that it makes no assumptions on the theory of (quantum) gravity that gives rise to the static and spherically symmetric metric (\ref{metrica adimensionale}). We only assume that it can (at least approximately) be cast into the form (\ref{EQT}) in an asymptotic regime for a physically reasonable $T_{\mu\nu}$. In particular we do not claim that the latter equation is the (complete) description of (quantum) gravity in all of space-time, but only describes static, spherically symmetric geometries at a sufficiently large distance from their rotational center. As we shall see, nevertheless physically plausible conditions on the form of $T_{\mu\nu}$ have imprints on the functions $f(z)$ and $h(z)$.

\item Our approach allows to interpret the functions $f(z)$ and $h(z)$ in (\ref{metrica adimensionale}) to encode quantum corrections to the (classical) Schwarzschild metric. In this case, $T_{\mu\nu}$ should rather have the interpretation of a vacuum expectation value in a suitable quantum state (see for example \cite{DEWITT1975295,Birrell:1982ix,Page:1982fm,Brown:1985ri}). 

\item {In the case of (\ref{EQT}) describing a rotationally symmetric black hole, we consider it to be static. In particular,} in the case of quantum black holes, which can evaporate by emitting Hawking radiation, we assume (\ref{EQT}) to hold in an (adiabatic) regime that approximates a static equilibrium state.

\end{itemize}
Under these conditions, the form of the metric (\ref{metrica adimensionale}) together with (\ref{PosMetricFunctions}) implies that $T_{\mu\nu}$ in (\ref{EQT}) is an energy-momentum tensor of type I (see appendix~\ref{App:Geometry}) in the classification of Ellis and Hawking \cite{Hawking:1973uf} (see appendix~\ref{App:EllisHawking}). Concretely, using the eigenvectors (\ref{Eigenvectors}), we can write the stress-energy tensor in the following form \cite{CosenzaHerrera,Cadoni:2022chn}
\begin{equation}
\label{emtensor}
    T_{\mu\nu}=(\epsilon+p_\perp)u_\mu u_\nu+ p_\perp g_{\mu\nu} -(p_\perp-p_ \parallel)w_\mu w_\nu\ .
\end{equation}
The three eigenvalues $\epsilon$, $p_\parallel$ and $p_\perp$ are called energy density, radial pressure and tangential pressure, respectively:
\begin{align}
&\epsilon=\frac{1-f-zf'}{8 \pi   z^2}\,,&&p_\parallel=\frac{f-1+zf\frac{h'}{h}}{8 \pi   z^2}\,,&&p_\perp=\frac{z}{2}p'_{\parallel}+\frac{zh'}{4h}(\epsilon+p_\parallel)+p_\parallel\,,\label{DefEpsP}
\end{align}
where primes denote derivatives with respect to $z$. 

\subsection{Physicality Conditions}\label{Sect:ConditionsT}
We next discuss physicality conditions for the metric function (\ref{metrica adimensionale}) and their interpretation in terms of the energy momentum tensor introduced through equation~(\ref{EQT}):
\begin{enumerate}
\item[\emph{(i)}] {\bf point-wise energy conditions:}\\
These are conditions imposed at a fixed point $P$ in space-time, which in the case of the spherically symmetric and static metric (\ref{metrica adimensionale}) means conditions at fixed $z$. They are motivated classically as conditions on the energy density that certain observers can measure locally. In this work we shall consider \cite{Hawking:1973uf,Visser:1996iw,Ford:1994bj,Martin-Moruno:2013wfa}
\begin{itemize}
\item weak energy condition (WEC): $T_{\mu\nu} v^\mu v^\nu\geq 0$ for any time-like vector $v^\mu$\\
This corresponds to the condition that the energy density measured by any time-like observer (\emph{i.e.} moving on a time-like curve) cannot be negative.
\item strong energy condition (SEC): $\left(T_{\mu\nu}-\frac{1}{2}\,{T^\rho}_\rho\, g_{_{\mu\nu}}\right)v^\mu v^\nu\geq 0$ for any time-like vector~$v^\mu$\\ 
Commonly (see \cite{Hawking:1973uf,Curiel:2014zba,Kontou:2020bta} for a discussion), this condition is interpreted geometrically in the form of the time-like convergence condition: indeed, the equivalent formulation $R_{\mu\nu}v^\mu v^\nu\geq 0$ implies that a congruence of time-like geodesics with vanishing rotation locally converges (see appendix~\ref{App:Raychaudhuri} for further explanations).
\item dominant energy condition (DEC): $T_{\mu\nu} v^\mu v^\nu\geq 0$ and $T^{\mu\nu}v_\nu$ is non-spacelike for any time-like vector $v^\mu$\\
This is equivalent to stating that any time-like observer finds a non-negative local energy density and the energy flow vector is non-spacelike 
\item null energy condition (NEC): $T_{\mu\nu} n^\mu n^\nu \geq 0$ for any null vector $n^\mu$. 
This corresponds to the condition that the energy density measured by any null observer (\emph{i.e.} moving on a null curve) cannot be negative.
\end{itemize}
These conditions are generally not independent but imply one-another according to the following scheme (see~\cite{Kontou:2020bta})
\begin{center}
\begin{tikzpicture}
\node at (0,0) {\fbox{\parbox{0.9cm}{DEC}}};
\draw[ultra thick,->] (0.7,0) -- (1.75,0);
\node at (2.5,0) {\fbox{\parbox{1cm}{WEC}}};
\draw[ultra thick,->] (3.2,0) -- (4.3,0);
\node at (5,0) {\fbox{\parbox{0.9cm}{NEC}}};
\draw[ultra thick,<-] (5.7,0) -- (6.8,0);
\node at (7.5,0) {\fbox{\parbox{0.825cm}{SEC}}};
\end{tikzpicture}
\end{center}

\noindent
The above conditions have been analysed and shown to hold for numerous classical systems. In particular in the case of GR, they are also linked to geometric properties, notably, through the Raychaudhouri equation (\ref{Raychaudhuritime}) and (\ref{Raychaudhurinull}), the geodesic motion of free-falling observers (see Appendix~\ref{App:Raychaudhuri} for more details). However, there are well known examples of quantum systems in which these conditions are violated \cite{Epstein:1965zza,Fewster:2002ne,Visser:1996iv} (see also the summary article \cite{Barcelo:2002bv}). Recent work has therefore focused on defining quantum-versions of these conditions, which put bounds on the amount to which they are allowed to be violated at the quantum level. Indeed, although negative energy densities (corresponding to violations of the above conditions) can lead to a violation of the second law of thermodynamics \cite{Ford:1978qya}, the latter can be avoided if the violation of the energy condition respects certain bounds \cite{Ford:1978qya,Ford:1990id}.\footnote{The bound in \cite{Ford:1978qya} was orignially defined for the energy flux, which we shall introduce as a condition later on.} The resulting in-equalities are often called \emph{quantum energy inequalities} (QI), \emph{e.g.} \cite{Ford:1994bj,Ford:1997fa,Kontou:2020bta} (see also the summary article \cite{Barcelo:2002bv}).

\item[\emph{(ii)}] {\bf averaged energy conditions:}\\
We can demand that the above conditions hold only when averaged over a certain region of space-time. The most common version of these conditions average energy densities along (causal) geodesics, potentially weighted by certain functions (see \cite{Kontou:2020bta,Curiel:2014zba} for further comments).  For example we have the: 
\begin{itemize}
\item averaged weak energy condition (AWEC): $\int T_{\mu\nu} v^\mu v^\nu d\tau\geq 0$ for any time-like geodesic $v^\mu$ (with proper time $\lambda$)
\item averaged null energy condition (ANEC): $\int T_{\mu\nu} n^\mu n^\nu d\lambda\geq 0$ for any null geodesic $n^\mu$ with affine parameter $\lambda$
\end{itemize}
Even in classical examples, the averaged conditions are generically weaker than their point-like counterparts discussed under point \emph{(i)}: in \cite{Fewster:2006ti,Fewster:2007ec} it was shown that the AWEC holds for certain cases of a non-minimally coupled Klein-Gordon field, while the WEC is violated. Generalisation of the averaged energy conditions that include lower bounds for their violation  at the quantum level are known as (averaged) quantum energy inequalities (see for example \cite{Roman:2004xm,Fewster:2003vg})

\item[\emph{(iii)}] {\bf flux energy conditions}\\
For a time-like observer with velocity $\mathfrak{v}^\mu$, the energy flux is defined by $F^\mu=-T^{\mu\nu}\mathfrak{v}_\nu$.

The classical flux energy condition (FEC) \cite{Abreu:2011fr,Martin-Moruno:2013sfa} states that this flux cannot be space-like, \emph{i.e.} $F^\mu F_\mu\leq 0$. For quantum systems, this condition is generalised (QFEC) by imposing that the flux is not 'excessively' spacelike \cite{Martin-Moruno:2013sfa}, \emph{i.e.}
\begin{align}
F^\mu F_\mu \leq \zeta\,\left(\frac{\hbar N}{L^4}\right)^2\,(\mathfrak{u}_\mu \mathfrak{v}^\mu)\,,
\end{align}
where $\mathfrak{u}_\mu$ is the system 4-velocity, while $N$ is the number of quantum fields and $L$ a typical length-scale of the problem and $\zeta$ a positive number of order of unity.

\end{enumerate}

\subsection{Equations of State and their Minimal Field Realisation}\label{fluid interpretation}
As we have mentioned before, the physical interpretation of the conditions discussed in the previous Subsection rely on identifying $T_{\mu\nu}$ in eq.~(\ref{EQT}) with an energy momentum tensor (or a suitable expectation value) of some physical system. For an energy-momentum tensor of type I in the classification of \cite{Hawking:1973uf} this system can in general be thought of as an \emph{anisotropic fluid}. The latter in turn can be realised through physical field configurations in various ways. In \cite{Boonserm:2015aqa} it was shown that a static and spherically symmetric anisotropic fluid can be generally described via a combination of a massless scalar field $\phi$, an electric-like field $E$ in addition to a perfect fluid that is characterised by the energy momentum tensor
\begin{align}
&{\widehat{T}^\mu}{}_\nu=\left(\begin{array}{cccc}\rho & 0 & 0 & 0 \\ 0 & p & 0 & 0 \\ 0 & 0 & p & 0 \\ 0 & 0 & 0 & p \end{array}\right)\,,&&\text{with} &&p=\omega(\rho)\,,\label{DefIdealFluid}
\end{align}
where $\rho$ is the energy density and $p$ the pressure component and the function $\omega(\rho)$ characterises the \emph{equation of state}.  
Concretely, the different components of $T_{\mu\nu}$ can be expressed as \cite{Boonserm:2015aqa}: 
\begin{align}
&\epsilon=\rho+\frac{1}{2}E^2+\frac{1}{2}\big(\grad{\phi}\big)^2\,,&& p_\parallel=p-\frac{1}{2}E^2+\frac{1}{2}\big(\grad{\phi}\big)^2\,,&&  p_\perp=p+\frac{1}{2}E^2-\frac{1}{2}\big(\grad{\phi}\big)^2\label{pperp1} \ . 
\end{align}
Inverting these equations is not unique, however, the following solution 
\begin{align}
    &p=\frac{1}{2}(p_\parallel+p_\perp)\,,&&\rho=\epsilon-\frac{1}{2}\abs{p_\parallel-p_\perp}\,,\label{rho1}\\
    &\big(\grad{\phi}\big)^2=\max \{p_\parallel-p_\perp,0\}\,, 
    &&E^2=\max\{p_\perp-p_\parallel,0\}\ . \label{elec}
\end{align}
is minimal \cite{Boonserm:2015aqa} in the sense of assuming the least number of fields:
\begin{itemize}
\item for $p_\parallel>p_\perp$, the energy-momentum tensor (\ref{EQT}) can be mimicked by a perfect fluid and a massless scalar field
\item for $p_\parallel<p_\perp$, the energy-momentum tensor (\ref{EQT}) can be mimicked by a perfect fluid and a spherically symmetric $U(1)$ gauge field with a non-vanishing electric component.
\end{itemize}
The perfect fluid is in both cases characterised by the equation of state. While this equation can take various different forms, examples for constant $\omega$ include \cite{Carroll:2004st}
\begin{enumerate}
\item[(a)] $\omega=0$, \emph{i.e.} $p=0$, which is realised by a dust-like pressureless fluid
\item[(b)] $\omega=\rho/3$, \emph{i.e.} $\rho=3p$, which is induced by (electro and magnetic)-like radiation
\item[(c)] $\omega=-\rho$, \emph{i.e.} $\rho=-p$, which can be realised via a  massive scalar field in which the kinetic term yields a subleading effect with respect to the mass term at large distances. We note that a cosmological constant term in the Einstein action would also lead to a perfect fluid with $\rho=-p$ which, however, would not vanish at large distances and therefore will not be considered here (because of our assumption to recover the Schwarzschild geometry in the asymptotically distant regime).
\end{enumerate} 
These examples are summarised in the Table~\ref{fluido1}. 
 
\begin{table}[h!]
\setlength{\arrayrulewidth}{0.8pt}
\centering
\begin{tabular}{l|c|c|c|}
\cline{2-4}
& \multirow{2}{*}{$\textcolor{rossoCP3}{p=\frac{1}{3}\rho}$} & \multirow{2}{*}{$\textcolor{rossoCP3}{p=-\rho}$} & \multirow{2}{*}{$\textcolor{rossoCP3}{p=0}$} \\
& & &  \\ \hline
\multirow{2}{*}{$\textcolor{rossoCP3}{\mathbf{p_\parallel<p_\perp}}$} & Electric background  +   & Electric background   + & Electric background   + \\ &  radiation & \textit{static} massive scalar   & pressureless dust \\ \hline
\multirow{2}{*}{$\textcolor{rossoCP3}{\mathbf{p_\parallel>p_\perp}}$} & massless scalar  + & massless scalar  + & massless scalar  +  \\ &  radiation  & \textit{static} massive scalar   & pressureless dust \end{tabular}
\caption{Examples of minimal particle physics realisation for a gravitating anisotropic fluid. The conditions $p_\parallel < p_\perp$ and $p_\parallel> p_\perp$ distinguish between different sources here identified as an electric-like background field and the massless scalar field, respectively. The columns provide examples of specific equations of states for the ideal fluid   $p=\omega \rho$: The radiation-like equation of state for $\omega=1/3$ , the $\omega=-1$  that stems, for example, from a massive scalar field whose kinetic term is neglected (in this sense the field is \textit{static}) and $\omega=0$ that emerges, for example, via pressureless dust.}
\label{fluido1}
\end{table}


\section{Energy Conditions}\label{Sect:EnergyConditionsGen}
In a first step, we shall keep the functions $h(z)$ and $f(z)$ in (\ref{metrica adimensionale}) as general as possible to evaluate the conditions introduced in Section~ \ref{Sect:ConditionsT}. Indeed, we consider general metric functions in (\ref{metrica adimensionale}) that asymptotically approach the Schwarzschild form, \emph{i.e.}
\begin{align}
&f(z)=1-\frac{2\chi}{z}\,\phi(z)\,,&&h(z)=1-\frac{2\chi}{z}\,\psi(z)\,,&&\forall z>\za\,,\label{MetricFctGen}
\end{align}
where $\phi,\psi:\,[\za,\infty)\to \mathbb{R}$ are continuous, (twice) differentiable functions of $z$ such that
\begin{align}
&\lim_{z\to \infty}\phi(z)=1=\lim_{z\to \infty}\psi(z)\,,&&\text{and} &&\lim_{z\to \infty}\frac{\dd \phi(z)}{\dd z}=0=\lim_{d\to \infty}\frac{\dd \psi(z)}{\dd z}\,.\label{LimitInf}
\end{align}
Here $\za>0$ is simply a value up to which the metric functions (\ref{MetricFctGen}) are defined. In the case of a black hole, $\za\geq \zh$, where $\zh$ denotes the position of the (external) horizon. We shall implement later in Section~\ref{Sect:AsymptoticExpansion} that these functions depend on $z$ only through a physical distance $d(z)$, concretely in the form of an asymptotic expansion that is valid for sufficiently large distances from the black hole. Currently, however, we will not make use of this property.
\subsection{Point-wise Energy Conditions}
In order to work out the WEC, DEC, NEC and SEC, we first require a set of time-like and null vectors. A useful basis in this regard are the eigenvectors of the Einstein tensor ${G^\mu}_\nu$ as outlined in eq.~(\ref{FormT}). We can then write the following set of time-like vectors
\begin{align}
&u^\mu\,,&&\tau_0^\mu(\beta)=u^\mu+\beta w^\mu\,,&&\tau_{1,2}^\mu(\beta)=u^\mu+\beta v_{1,2}^\mu\,,&&\forall \beta\in(-1,1)\,.\label{SetTimeLikeVectors}
\end{align}
A useful set of null vectors is given by $\tau_{0,1,2}^\mu(\beta=\pm1)$.

\subsubsection{NEC}
We start by discussing the NEC, which is implied by other point-like energy conditions as discussed in the previous Section. Evaluating the NEC condition for the null vectors $\tau^{\mu}_{0,1,2}(\beta=\pm 1)$, we obtain
\begin{align}
&\mathfrak{c}_1=\epsilon+p_\parallel\geq 0 \ ,&&\text{and} &&\mathfrak{c}_2=\epsilon+p_\perp\geq 0  \ .\label{ec1}
\end{align} 
Expressing the eigenvalues of the Einstein tensor in terms of the functions $f$ and $h$ (see eq.~(\ref{FormT})), the condition $\mathfrak{c}_1$ in (\ref{ec1}) becomes
\begin{align}
\frac{1}{8\pi z}\,\left(f\,\frac{h'}{h}-f'\right)\geq 0\,.\label{NonTrivialC1}
\end{align}
Assuming that $f(z)>0$ for $z>\za$\footnote{In the case of a black hole, this is indeed the case outside of the horizon}, this relation can in fact be turned into the following integral inequality
\begin{align}
&\int_z^\infty dz'\,\frac{h'(z')}{h(z')}\geq \int_z^\infty dz'\,\frac{f'(z')}{f(z')}\,,&&\text{such that} &&\log(h(z))\leq \log(f(z))\,,&&\forall z>\za\,.
\end{align}
Here we have used the limit (\ref{LimitInf}). Furthermore, since for $x>0$ the function $\log(x)$ is monotonically growing, we have
\begin{align}
&h(z)\leq f(z)\,,&&\text{and} &&\psi(z)\geq \phi(z)\,,&&\forall z>\za\,.\label{RelhfGen}
\end{align}
The second condition $\mathfrak{c}_2$ in (\ref{ec1}) is more complicated and does not readily lead to simple conditions for the functions $\phi$ and $\psi$. We shall discuss it in more detail in Section~\ref{Sect:AsymptoticExpansion} by assuming an asymptotic expansion of the latter that is compatible with diffeomorphism invariance of the metric. 

Here we remark that the condition $\mathfrak{c}_2$ becomes significantly simpler in the special case $f=h$. Indeed, this case is allowed by (\ref{RelhfGen}) (in fact $\mathfrak{c}_1=0$ in this case) and the eigenvalues of ${T^\mu}_\nu$ take the following simple form
\begin{align}
\epsilon=-p_{||}=\frac{1-f-z f'}{8\pi z^2}=\frac{\chi}{4\pi z^2}\,\frac{\dd\phi}{\dd z}\,,&&p_{\perp}=\frac{2f'+zf''}{16\pi z}=-\frac{z}{2}\,\epsilon'-\epsilon\,,\label{EpsPDiff}
\end{align}
such that $\mathfrak{c}_2$ becomes
\begin{align}
&\mathfrak{c}_2=-\frac{z}{2}\,\epsilon'\geq 0\,,&&\forall z>z_H\,.
\end{align}
Since we assume $z>\za>0$, we can integrate this condition
\begin{align}
&0\geq \int_z^\infty dz'\,\epsilon'(z')=\lim_{z'\to \infty}\epsilon(z')-\epsilon(z)=-\epsilon(z)\,,&&\forall z>\za\,.\label{equivWEC}
\end{align}
Furthermore, using (\ref{EpsPDiff}), the condition $\epsilon(z)\geq 0$ can be integrated a second time
\begin{align}
&0\leq \int_z^\infty\frac{\dd \phi}{\dd z}=1-\phi(z)\,,&&\forall z> \za\,.
\end{align}
We can therefore summarise the relations in the form
\begin{align}
&\phi(z)=\psi(z)\leq 1\,,&&\text{and} &&\frac{d\phi}{dz}\geq 0\,,&&\forall z>\za\,,\label{NECsimple}
\end{align}
In the case of a black hole, with (outer) horizon $\zh$, and with $\za=\zh$, this means that the

\begin{wrapfigure}{l}{0.40\textwidth}
\parbox{6cm}{\begin{tikzpicture}
\draw[->] (0,0) -- (5,0);
\node at (5.3,0) {$z$};
\draw[->] (0,0) -- (0,2);
\node at (0,2.3) {$\phi(z)$};
\draw[dashed] (0,1.5) -- (5,1.5);
\node at (-0.2,1.5) {$1$};
\draw[dashed] (0,0.4) -- (5,0.4);
\node at (-0.25,0.4) {$\tfrac{\za}{2\chi}$};
\draw[dashed] (1,0) -- (1,2);
\node at (1,-0.25) {$\za$};
\draw[thick,red] (1,0.4) to [out=10,in=200] (2.5,0.8) to [out=20,in=180] (4.8,1.5) to [out=0,in=180] (5,1.5);
\end{tikzpicture}
\caption{Schematic drawing of the function $\phi$ compatible with eq.~(\ref{NECsimple})}
\label{Fig:SchemFuncPhi1}}
\end{wrapfigure}
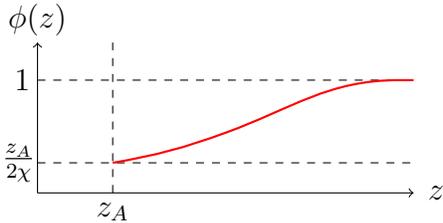

\noindent
function $\phi$ smoothly interpolates between the value $\frac{\zh}{2\chi}$ at the position of the horizon $\zh$ and $1$ for $z\to \infty$ as schematically shown in Figure~\ref{Fig:SchemFuncPhi1}. Notice that the function $\phi$ may have a saddle point for some $z\geq \zh$ (such that $\frac{d^2 \phi}{dz^2}=0=\frac{d \phi}{dz}$) but may not have a local extremum for $z>\zh$. Notice furthermore that positive definiteness of the first derivative of $\phi$ implies that $\zh\leq 2\chi$, \emph{i.e.} the position of the horizon is smaller than in the classical case.

\noindent
Still in the case of $f=h$, we can also consider the anisotropy parameter, which determines what type of classical field configuration can mimic the energy momentum tensor. Indeed, following the discussion in Section~\ref{fluid interpretation}, we find
\begin{align}
p_\perp-p_{||}=-\frac{z}{2}\,\epsilon'\geq 0\,.\label{Anisotropy}
\end{align}
The isotropic case $p_\perp=p_{||}$ is equivalent to $\frac{d^2\phi}{dz^2}=0$ which has as only solution compatible with the condition (\ref{LimitInf}) $\phi(z)=1$ with $\za=1$, which, however, is the Schwarzschild metric. For $p_\perp\neq p_{||}$ , eq.~(\ref{Anisotropy}) implies that the energy momentum tensor associated with such a metric can always be mimicked by an electric field and a perfect fluid (see Section~\ref{fluid interpretation}). For the latter we can derive the equation of state
\begin{align}
\rho=\epsilon-\frac{1}{2}|p_{||}-p_\perp|=\epsilon-\frac{1}{2}(p_\perp-p_{||})=-\frac{1}{2}(p_{||}+p_\perp)=-p\,.\label{EoSGenfh}
\end{align}

\subsubsection{WEC}
Projecting the energy-momentum tensor (\ref{emtensor}) with the time-like vectors (\ref{SetTimeLikeVectors}) gives rise to the following conditions
\begin{align}
&\epsilon\geq 0\,,&&\qq{and} &&\epsilon+\beta^2 p_{||}\geq 0\,,&&\qq{and} &&\epsilon+\beta^2 p_\perp\geq 0\,,&&\forall \beta\in(-1,1)\,,\label{WECtang}
\end{align}
which are usually written in the form \cite{Curiel:2014zba}
\begin{align}
&\epsilon\geq 0\,,&&\qq{and}&&\epsilon+p_{||}\geq0\,,&&\qq{and} &&\epsilon+p_\perp\geq 0\,.\label{WEC}
\end{align} 
The last two of these relations are in fact the NEC-conditions (\ref{ec1}), the first of which we have already shown above to lead for the metric (\ref{MetricFctGen}) to the condition (\ref{NECsimple}). Using the expression for $\epsilon$ in (\ref{DefEpsP}) in terms of the function $f$, along with (\ref{MetricFctGen}), the first relation in (\ref{WEC}) becomes
\begin{align}
\frac{\chi}{4\pi z^2}\,\frac{d\phi}{dz}\geq 0\,,&&\forall z>\za\,,\label{DerPhi}
\end{align}
imposing that the first derivative of $\phi$ (as a function of $z$) needs to be positive. Integrating this relation and using (\ref{LimitInf}) this furthermore implies
\begin{align}
0\leq \int_z^\infty dz'\,\frac{d\phi(z')}{dz'}=1-\phi(z)\,.\label{PhiIneq}
\end{align}
The first two equations in (\ref{WEC}) can therefore be summarised as
\begin{align}
&\psi(z)\geq \phi(z)\,,&&\text{and} && 1\geq \phi(z)\,,&&\text{and} &&\frac{d\phi}{dz}\geq 0\,, &&\forall z>\za\,.\label{RelhfGen1}
\end{align}
As in the case of the NEC, the last equation in (\ref{WEC}) is rather involved and we shall discuss it in more detail in the following Section, assuming a particular asymptotic series expansion of the functions $\phi$ and $\psi$. We remark, however, that for the particular case $f=h$, we have shown in the previous Subsection that $\mathfrak{c}_2$ leads to (\ref{equivWEC}), which in fact the first equation in (\ref{WEC}). In this particular case therefore, the WEC and the NEC are completely equivalent.

\subsubsection{DEC}
We next consider the DEC, using the time-like vectors (\ref{SetTimeLikeVectors}). In addition to the relations (\ref{WECtang}) these also lead to the conditions
\begin{align}
&(\epsilon+\beta\, p_{||})(\epsilon-\beta\, p_{||})\geq 0\,,&&\qq{and} &&(\epsilon+\beta\, p_{\perp})(\epsilon-\beta\,p_{\perp})\geq 0\,.
\end{align}
The DEC is therefore commonly formulated as \cite{Curiel:2014zba}
\begin{align}
&\epsilon\geq 0\,,&&\qq{and} &&\epsilon\geq |p_{||}|\,,&&\qq{and}&&\epsilon\geq |p_\perp|\,,\label{DEC}
\end{align}
(which implies the condition (\ref{WEC})). The first condition implies (\ref{DerPhi}) and (\ref{PhiIneq}), just as in the case of the WEC. The second equation in (\ref{DEC}) can be rewritten in as
\begin{align}
&\left\{\begin{array}{l}1-f-zf\,h'/h\geq 0\,,\\ h'/h\geq f'/f\,.\end{array}\right.&&\text{or} &&\left\{\begin{array}{l}1-f-zf\,h'/h\leq 0\,, \\ 2-2f-zf'\geq zf\,h'/h\,.\end{array}\right.
\end{align}
Both inequalities imply $h'/h\geq f'/f$, such that the the DEC indeed implies the same relations (\ref{RelhfGen}) we had also found previously for the WEC. However, a more detailed analysis of (\ref{DEC}) is rather involved and we shall provide more details in the following Section, assuming an asymptotic expansion of the function $\phi$ and $\psi$. We remark, however, that in the particular case $f=h$, in addition to (\ref{NECsimple}), the third equation in (\ref{DEC}) implies
\begin{align}
\label{DEC36}
&\phi'(z)\geq \frac{1}{2}\,|z\,\phi''(z)|\,.
\end{align}

 \subsubsection{SEC}
Using the time-like vectors (\ref{SetTimeLikeVectors}), the conditions for the SEC can be written in the form
\begin{align}
&\epsilon+p_{||}+2 p_\perp\geq0\,,&&\text{and} &&\epsilon+p_{||}+2p_\perp+\beta^2\left(\epsilon+p_{||}-2p_\perp\right)\geq 0\,,\nonumber\\
& &&\text{and} &&\epsilon+p_{||}+2p_\perp+\beta^2\left(\epsilon-p_{||}\right)\geq 0\,,
\end{align}
which are commonly presented in the form \cite{Curiel:2014zba}
\begin{align}
&\epsilon+p_{||}+2 p_\perp\geq0\,,&&\text{and} &&\epsilon+p_{||}\geq 0\,,&&\text{and} &&\epsilon+p_\perp\geq 0\,.\label{SEC}
\end{align}
The last two inequalities are in fact the NEC conditions, implying therefore (\ref{RelhfGen}), in agreement

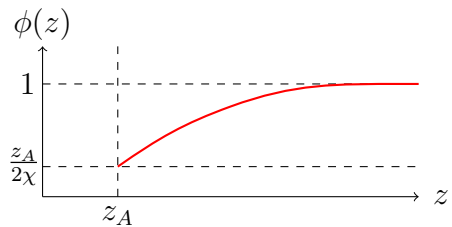
\begin{wrapfigure}{r}{0.40\textwidth}
\parbox{6cm}{\begin{tikzpicture}
\draw[->] (0,0) -- (5,0);
\node at (5.3,0) {$z$};
\draw[->] (0,0) -- (0,2);
\node at (0,2.3) {$\phi(z)$};
\draw[dashed] (0,1.5) -- (5,1.5);
\node at (-0.2,1.5) {$1$};
\draw[dashed] (0,0.4) -- (5,0.4);
\node at (-0.25,0.4) {$\tfrac{\za}{2\chi}$};
\draw[dashed] (1,0) -- (1,2);
\node at (1,-0.25) {$\za$};
\draw[thick,red] (1,0.4) to [out=35,in=200] (2.5,1.2) to [out=20,in=180] (4.8,1.5) to [out=0,in=180] (5,1.5);
\end{tikzpicture}
\caption{Schematic drawing of the function $\phi$ compatible with eq.~(\ref{NECsimple}) and eq.~(\ref{ineqPhi}).}
\label{Fig:SchemFuncPhiSEC}}
\end{wrapfigure}

\noindent
to the discussion in Section~\ref{Sect:ConditionsT}. As for the remaining  conditions, an analysis of the full system shall be relegated to the next Section for an asymptotic expansion of the functions $f$ and $h$. 

\noindent
Here we only remark that in the particular case $f=h$, in addition to eq.~(\ref{NECsimple}), the first equation of (\ref{SEC}) leads to the inequality
\begin{equation}
  -\frac{\chi  \phi ''(z)}{4 \pi  z} \geq 0\,.\label{ineqPhi}
\end{equation}
Therefore the second derivative of $\phi(z)$ must always be negative. A schematic drawing of such a function is shown in Figure~\ref{Fig:SchemFuncPhiSEC}. We note, however, that in principle the function $\phi$ may still have saddlepoints.

\subsection{Averaged Energy Conditions}
For completeness, we shall also discuss the averaged weak- and null energy condition in this Section. As we mentioned before, while in principle averages over generic regions of space-time can be considered, the most common versions of the AWEC and ANEC use entire (causal) geodesics. Since, however, eq.~(\ref{EQT}) was assumed to only hold in an asymptotic region (sufficiently far from the rotational center of the geometry) also the energy momentum tensor can only be reasonably defined for some $z>\za$. Therefore, the average of the energy conditions can also only be defined over regions with $z>\za$ and $\za$ will enter explicitly. Since the latter is not necessarily a physical parameter (but rather a limitation of the description in eq.~(\ref{EQT})), the physical conclusions drawn from these averaged conditions are therefore in general somewhat limited. In the following we shall therefore outline only two examples of averaged conditions.
\begin{itemize}
\item AWEC: 
Using for example the radial time-like geodesic in  (\ref{TimeLikeGeo}) for $z\in[\za,\infty)$ we can write the averaged WEC in the following form 
\begin{align}
&0\leq \int_{\za}^\infty dz\,\frac{h((1-f)h-z(1+\beta^2)f')+z fh'(1+\beta^2-h)}{8\pi z^2 h \sqrt{fh}\sqrt{1+\beta^2-h}}\,,&&\forall \beta\in\mathbb{R}\text{ with }\beta^2<1\,.\label{AWECcond}
\end{align}
\item ANEC: Using for example the radial null geodesic in (\ref{NullGeo}) for $z\in[\za,\infty)$, we can average the NEC condition in the following manner
\begin{align}
0\leq \int_{\za}^\infty dz\,\frac{f(z)\,h'(z)-h(z)\,f'(z)}{8\pi z h(z)\,\sqrt{f(z)\,h(z)}}=-\int_{\za}^\infty \frac{dz}{4\pi z}\,\frac{d}{dz}\left(\sqrt{\frac{f(z)}{h(z)}}\right)\,.\label{ANECcond}
\end{align}
After partial integration, this equation becomes
\begin{align}
\frac{1}{4\pi \zh}\,\sqrt{\frac{f(\zh)}{h(\zh)}}\ \geq \int_{\za}^\infty \frac{dz}{4\pi z^2}\,\sqrt{\frac{f(z)}{h(z)}}\,.
\end{align}
In the case $f=h$, this relation is indeed trivially satisfied. 
\end{itemize}

\subsection{Flux Energy Condition}\label{Sect:RelFEC}
Using the time-like vectors $u^\mu$ and $\tau_{0,1,2}^\mu(\beta)$ in (\ref{SetTimeLikeVectors}), we can write the flux energy conditions in the form
\begin{align}
&\epsilon^2\geq 0\,,&&\text{and} &&\epsilon^2\geq \beta^2\, p^2_{||}\,,&&\text{and} &&\epsilon^2\geq \beta^2\,p_\perp^2\,,&&\forall \beta\in(-1,1)\,,
\end{align}
which can be formulated as \cite{Abreu:2011fr,Martin-Moruno:2013sfa}
\begin{align}
&\epsilon^2\geq p^2_{||}\,,&&\text{and} &&\epsilon^2\geq p_\perp^2\,.\label{FEC}
\end{align}
Although rather involved, we can analyse these conditions for the particular case of the metric function (\ref{MetricFctGen}): the first condition in (\ref{FEC}) can be written in the form
\begin{align}
\frac{f}{64\pi^2 z^3 h^3}\left(\frac{f'}{f}-\frac{h'}{h}\right)\left(\frac{2f-2+zf'}{f}+z\frac{h'}{h}\right)\geq 0\,,
\end{align}
which can be resolved in two different ways
\begin{align}
&\left\{\begin{array}{l}\frac{f'}{f}\geq \frac{h'}{h}\text{  and}\\[4pt] \frac{2f-2+zf'}{f}+z\,\frac{h'}{h}\geq 0\end{array}\right.&&\text{or} &&\left\{\begin{array}{l}\frac{f'}{f}\leq \frac{h'}{h}\text{  and}\\[4pt] \frac{2f-2+zf'}{f}+z\,\frac{h'}{h}\leq 0\end{array}\right.
\end{align}
Inserting the first of these condition in the later, we can deduce
\begin{align}
&2f-2+zf'+zf'=-4\chi \phi'\geq 0\,,&&\text{or} &&2f-2+zf'+zf'=-4\chi \phi'\leq 0\,,
\end{align}
which entails that $\phi$ is a monotonic function which cannot change its sign. The same conclusion can also be drawn from the second condition in (\ref{FEC}). A more general analysis of (\ref{FEC}) is relegated to Section~\ref{Sect:AsymptoticExpansion}, where we shall assume an asymptotic expansions of the functions $\phi$ and $\psi$, which simplifies the discussion. 

We close this subsection by remarking that even for general $\phi$ and $\psi$ in (\ref{MetricFctGen}), (\ref{FEC}) simplifies if we assume $f=h$. Indeed, in this case the first condition in (\ref{FEC}) is trivially satisfied, while the second condition can be re-written in the form
\begin{align}
\frac{(2-2f+z^2 f'')(2-2f-4z f'-z^2 f'')}{256\pi^2 z^4 }&=\frac{\chi^2}{64\pi^2 z^4}\left[(2\phi')^2-(z\phi'')^2\right]\nonumber\\
&=-\frac{4\chi^2}{256\pi^2 z^2}\,\left(\frac{\phi'}{z^2}\right)'\,(z^2\,\phi')'\geq 0\,.\label{CondFlux}
\end{align}
This implies that $|2\phi'|\geq |z\phi''|$ which can be turned into a non-trivial statement about how the function $\phi$ behaves for large $z$. Indeed, the second line in (\ref{CondFlux}) implies that either $(z^2 \phi')'=0$ or $(z^2 \phi')'$ and $(\phi'/z^2)'$ have opposite signs. The $|2\phi'|\geq |z\phi''|$ condition is similar to the DEC in \eqref{DEC36}. To analyse the latter case in the asymptotic regime, \emph{i.e.} far away from the rotational center such that the sign of $\phi'$ remains constant, let $\mathfrak{s}=\text{sign}(\phi')$ for $z\gg 1$, then the only possible combination of signs for these conditions can be integrated as follows
\begin{align}
&\mathfrak{s}\,\int_z^\infty dz'\left(\phi'/z^{\prime 2}\right)'=-\mathfrak{s}\frac{\phi'}{z^2}\leq 0\,,&&\text{and} &&\mathfrak{s}\,\int_z^\infty dz'\left(z^{\prime 2}\,\phi'\right)'=\mathfrak{s}\lim_{z\to \infty}(z^2\,\phi')-\mathfrak{s}\,z^2\,\phi'\geq 0\,.
\end{align}
Among these, the first relation is always satisfied, whereas the second condition requires that $\lim_{z\to \infty}(z^2\,\phi')\neq0$ to be compatible, \emph{i.e.} $\phi'$ cannot tend to zero faster than $1/z^2$: if $\mathfrak{s}\,\lim_{z\to \infty}(z^2\,\phi')\to \infty$, the second condition is also trivially satisfied (and we find no further information on $\phi'$). If $\lim_{z\to \infty}(z^2\,\phi')$ is finite (\emph{i.e.} $\phi'(z)\sim \frac{\mathfrak{s}c}{z^2}+\mathfrak{o}(z^{-2})$, for $c\in\mathbb{R}_+$), the sign of the subleading correction needs to be $-\mathfrak{s}$ in the asymptotic regime. We shall discuss this condition further in the following Section.





\section{Conditions for Asymptotic Expansions}\label{Sect:AsymptoticExpansion}
The form of the metric functions (\ref{MetricFctGen}) was motivated by demanding to reproduce asymptotically the Schwarzschild geometry, which only requires the limits in eq.~(\ref{LimitInf}). As we have seen in the previous Section, various energy conditions (which are motivated from eq.~(\ref{EQT}) for an energy-momentum tensor $T_{\mu\nu}$ that satisfies basic physicality conditions) already impose non-trivial requirements for the functions $(f,h)$ (or equivalently $(\phi,\psi)$). These can be made more stringent by using the form advocated in \cite{Bonanno:2000ep, Hayward_2006,Platania:2019kyx,Knorr:2022kqp,Binetti:2022xdi,Eichhorn:2022bgu,bardeen1968proceedings}: indeed, demanding diffeomorphism invariance of the metric, it was proposed to write $(\phi,\psi)$ as functions of an invariant (physical) quantities. Concretely, for the latter, the physical distance measured from the rotational center was chosen.

As remarked in \cite{Bonanno:2000ep}, the concrete choice of this distance is ambiguous, however, at the same time physical quantities do not depend on it. For example, in appendix~\ref{App:SchemeDependence} we argue that shifting this distance by a ($z$-independent) constant can be accommodated in our approach below. In this Section we use this freedom to define the physical distance $d(z)$ in a way that differs from \cite{Bonanno:2000ep,Binetti:2022xdi} by a(n additive) constant, namely through the following differential equation with an asymptotic boundary condition
\begin{align}
&\frac{d}{dz}\,d=\frac{1}{\sqrt{|f(z)|}}\,,&&\text{with} &&\lim_{z\to \infty} (d(z)-d_0(z))=0\,.\label{DefDistanceAsymp}
\end{align}
Here $d_0$ is the distance from the origin in the (classical) Schwarzschild metric with mass parameter $\chi$, \emph{i.e.}
\begin{align}
&d_0(z)=\pi \chi+2\chi\tanh^{-1}\sqrt{1-\frac{2\chi}{z}}+\sqrt{z(z-2\chi)}\,,&&\forall z>2\chi\,.
\end{align}
Furthermore, to simplify the analysis of the different energy conditions, we consider an asymptotic expansion of $(f,h)$
\begin{align}
&f(z)=1-\frac{2\chi}{z}\left(1+\sum_{n=1}^\infty\ \frac{\omega_n}{d(z)^{n}}\right)\,,&&\text{and} &&  h(z)=1-\frac{2\chi}{z}\left(1+\sum_{n=1}^\infty\ \frac{\gamma_n}{d(z)^{n}}\right) \ , \label{Functionsfh}
\end{align}
where $\omega_n,\gamma_n$ are effective coefficients encoding the deformation from Einstein gravity (and the deviation from the Schwarzschild metric). While the motivation for these expansions is the same as in \cite{Binetti:2022xdi}, they technically generalise the discussion there by also including odd powers of the inverse distance $u=1/d(z)$.\footnote{We recover the metric and results of \cite{Binetti:2022xdi} when taking  $\gamma_n=\omega_n$ and restricting  $n$ to be even.}

\begin{figure}[h]
\centering
\scalebox{0.96}{\parbox{13.1cm}{\begin{tikzpicture}[scale = 1.50]
\draw[fill=black] (0,0) circle (0.075);
\draw[->] (0,0) -- (7,0);
\draw[red,thick] (4.5,0) -- (7,0);
\node at (7.65,0.15) {$z\,,d\to \infty$};
\node at (7.45,-0.15) {$u\to 0$};
\draw[ultra thick,domain=-20:20,dashed] plot ({1.5*cos(\x)},{1.5*sin(\x)}); 
\node at (0,-0.3) {$z=0$}; 
\node at (1.5,0.75) {horizon}; 
\node at (1.5,-0.75) {$z_H\,,d_H\,,\frac{1}{u_H}$}; 
\draw[fill=black] (2.75,0) circle (0.05);
\node at (2.75,0.3) {$\za\,,d_A\,,\frac{1}{u_A}$}; 
\draw[fill=black] (4.5,0) circle (0.05);
\node at (4.5,-0.3) {$z_c\,,d_c\,,\frac{1}{u_c}$}; 
\node[red] at (5.75,0.55) {range of validity}; 
\node[red] at (5.75,0.25) {of (\ref{Functionsfh})}; 
\end{tikzpicture}
}}
\caption{\sl Schematic representation of the range of validity of the series expansions (\ref{DefSeriesExpansion}), which defines the scale $d_c=1/u_c$. We have introduced the variable $u(z)=1/d(z)$. {Furthermore, we have also included the coordinate $\za$ (along with the distance $d_A=d(\za)$ and $u_A=1/d_A$) beyond which we assume the Einstein equation (\ref{EQT}) to hold, as explained in Section~\ref{Sect:EnergyMomentum}.} Finally, in case of the metric representing a black hole, the figure also shows its position relative to the range of validity of~(\ref{Functionsfh}).}
\label{Fig:ScaleBH}
\end{figure}
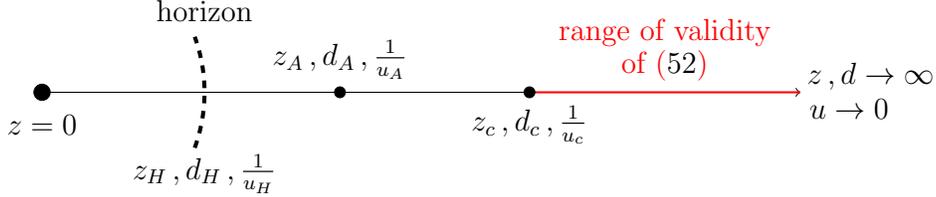

An important aspect of the series (\ref{Functionsfh}) is their range of validity: indeed, while an expansion in inverse powers of $d$ guarantees that asymptotically the Schwarzschild metric is recovered\footnote{Notice also, since $d'(z)=\frac{1}{\sqrt{|f(z)|}}$, which tends to $1$ for $z\to \infty$, the first derivatives of these functions vanish asymptotically, which is compatible with (\ref{LimitInf}).}, in general, such an expansion converges (against the actual metric functions) only for sufficiently large $d(z)$. This is schematically shown in Figure~\ref{Fig:ScaleBH}: there exists a critical distance $d_c$ (and thus also a critical inverse distance $u_c$ and a critical $z_c$) below which the series (\ref{Functionsfh}) no longer (correctly) represent the metric functions $f$ and $h$. Concretely, this scale can be the radius of convergence of the series (\ref{Functionsfh}). {In the following we shall assume $z_c\geq \za$, such that (\ref{EQT}) is valid in the entire region of validity of (\ref{Functionsfh}),} In the case of a black hole metric, as we shall discuss in more detail in Section~\ref{Sect:RadiusConvergence}, while the latter can be of the order of the (outer) black hole horizon (\emph{i.e.} $d_c\sim d_A\sim d_H$), such that (\ref{Functionsfh}) is valid in the entire space-time outside the black hole horizon, this requires a priori that the coefficients $(\omega_n,\gamma_n)$ depend in a specific fashion on the mass of the black hole. On the other hand, as we shall discuss in Section~\ref{Sect:CordConstraints}, the existence of the scale $d_c$ in conjunction with the energy conditions requires certain non-trivial relations among the coefficients $(\omega_n,\gamma_n)$. 

Our starting point in the following, however, is  the asymptotic regime (\emph{i.e.} large $z$ or, equivalently, large $d$) in which the metric is near flat (and $d(z)$ becomes proportional to $z$ itself to leading order, as follows from (\ref{DefDistanceAsymp})). Notice in this regard also that the definition (\ref{DefDistanceAsymp}) is self-consistent in the sense that it depends on the function $f$ only in the region where its series expansion (\ref{Functionsfh}) is defined (\emph{i.e.} the region marked red in Figure~\ref{Fig:ScaleBH}). Using the distance measured from the center of the black hole (as for example in \cite{Bonanno:2000ep,Binetti:2022xdi}) would require additional information about the function $f$ outside of this region.


\subsection{Conditions to Leading Order}\label{NECLeadingOrder}
As explained in more detail in Appendix~\ref{App:Distance}, in the asymptotic region, we can solve (\ref{DefDistanceAsymp}) order by order in an expansion in $z$ (see eq.~(\ref{DefSeriesExpansion})). The latter can also be inverted to express $d$ order by order in $z$ (see eq.(\ref{Expansionzd})). Using these expressions, we can calculate to leading order
\begin{align}
&\epsilon=-\frac{\chi \omega_1}{4\pi z^4}+\mathfrak{o}(z^{-4})\,,&&p_{||}=\frac{\chi(2\gamma_1-\omega_1)}{4\pi z^4}+\mathfrak{o}(z^{-4})\,,&&p_\perp=\frac{\chi(\omega_1-2\gamma_1)}{4\pi z^4}+\mathfrak{o}(z^{-4})\,.\label{SerExpLeadingEpsP}
\end{align}
To analyse in more detail the various energy conditions discussed in Section~\ref{Sect:ConditionsT}, we first consider the case $(\omega_1,\gamma_1)\neq (0,0)$, which we shall relax in the subsequent Subsubsection.

\subsubsection{Case $(\omega_1,\gamma_1)\neq(0,0)$}
We first restrict ourselves to the case $(\omega_1,\gamma_1)\neq(0,0)$. Under this assumption, we start by discussing the point-wise energy conditions outlined in point \emph{(i)} of Section~\ref{Sect:ConditionsT}. Inserting the expansions (\ref{SerExpLeadingEpsP}) into the conditions (\ref{DEC}), (\ref{WEC}), (\ref{ec1}) and (\ref{SEC}) respectively, they all reduce for $(\omega_1,\gamma_1)\neq (0,0)$ to 
\begin{align}
\text{DEC, WEC, NEC, SEC:}\hspace{1cm}\omega_1\lesssim\gamma_1<0\,.\label{LeadingDWNSEC}
\end{align}
A few comments are in order concerning this result
\begin{itemize}
\item The notation $\omega_1\lesssim\gamma_1$ means that the energy conditions are always satisfied for $\omega_1<\gamma_1<0$ (for large $z$), while for $\omega_1=\gamma_1<0$ eqs.~(\ref{DEC}), (\ref{WEC}), (\ref{ec1}) and (\ref{SEC}) involve conditions for coefficients $(\omega_k,\gamma_k)$ with $k>1$. Concretely, for $\omega_1=\gamma_1<0$, the leading $\mathcal{O}(z^{-4})$ terms are all either compatible with the energy conditions in a non-trivial fashion or vanish. In the latter case, we need to inspect the subleading $\mathcal{O}(z^{-n})$ contributions (for $n>4$): empirically, we have observed up to $n=9$ that they are tantamount to
\begin{align}
&\text{DEC, WEC, NEC, SEC:}\nonumber\\
&\lim_{z\to\infty} z^{3+n}\,(\epsilon+p_{||})=\frac{(n+1)\chi}{4\pi}\,(\gamma_n-\omega_n)\geq 0\,,&&\text{if} &&\gamma_i=\omega_i\hspace{0.3cm}\forall ~ 1\leq i<n\,,
\end{align}
which requires $\gamma_n>\omega_n$ for the first coefficients $(\gamma_n,\omega_n)$ that are different from one another.\footnote{The case $\omega_n=\gamma_n$ $\forall n\geq 1$ (\emph{i.e.} $f(z)=h(z)$), is briefly discussed in Subsection~\ref{f=h}.}

\item For $\gamma_1=0$ (but $\omega_1<0$), the leading order contributions $\mathcal{O}(z^{-4})$ in eqs.~(\ref{DEC}), (\ref{WEC}), (\ref{ec1}) and (\ref{SEC}) are all either satisfied non-trivially or vanish. In the latter case, however, the subleading logarithmically enhanced terms of the order $\frac{\log z}{z^5}$ are incomaptible with $\omega_1<0$, for example
\begin{align}
\epsilon+p_{||}\big|_{\gamma_1=0}=\frac{\chi^2\omega_1}{8\pi z^5}\,\log z+\frac{\chi}{8\pi z^5}\left[\omega_1\chi(\pi-2+\log(2/\chi))-\omega_2-9\gamma_2\right]+\mathfrak{o}(z^{-5})\,.
\end{align}
\item The condition (\ref{LeadingDWNSEC}) is compatible with (\ref{RelhfGen}), which we have derived on general grounds for all point-wise energy conditions.
\end{itemize}

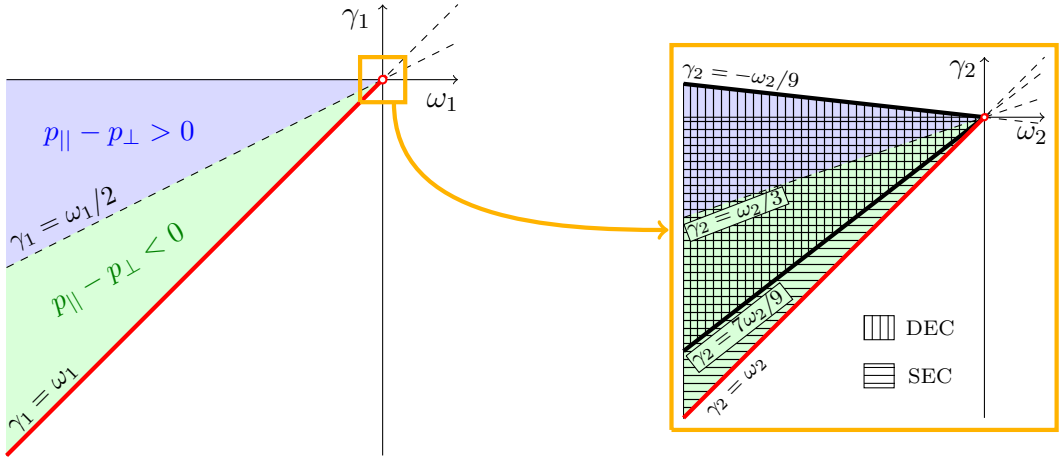
\begin{figure}[h]
\begin{center}
\begin{tikzpicture}
\draw[blue!15!white,fill=blue!15!white] (-5,0) -- (0,0) -- (-5,-2.5) -- (-5,0);
\draw[green!15!white,fill=green!15!white] (-5,-2.5) -- (0,0) -- (-5,-5) -- (-5,-2.5);
\draw[->] (-5,0) -- (1,0);
\node at (0.8,-0.3) {$\omega_1$};
\draw[->] (0,-5) -- (0,1);
\node at (-0.35,0.8) {$\gamma_1$};
\draw[dashed] (-5,-5) -- (1,1);
\node[rotate=45] at (-4.5,-4.25) {\footnotesize$\gamma_1=\omega_1$};
\draw[dashed] (-5,-2.5) -- (1,0.5);
\node[rotate=25] at (-4.25,-1.85) {\footnotesize$\gamma_1=\omega_1/2$};
\draw[thick,red, ultra thick] (-5,-5) -- (0.05,0.05);
\draw[thick,red,fill=white] (0,0) circle (0.05cm);
\node[rotate=0,blue] at (-3.5,-0.75) {\small$p_{||}-p_\perp>0$};
\node[rotate=30,green!50!black] at (-3.5,-2.55) {\small$p_{||}-p_\perp<0$};
\draw[red!30!yellow,ultra thick] (-0.3,-0.3) -- (0.3,-0.3) -- (0.3,0.3) -- (-0.3,0.3) -- (-0.3,-0.3);
\draw[red!30!yellow,ultra thick,->] (0.15,-0.3) to [out=270,in=180] (3.8,-2); 
\begin{scope}[xshift=8cm,yshift=-0.5cm, scale=0.8]
\draw[red!30!yellow,ultra thick] (-5.2,-5.2) -- (-5.2,1.2) -- (1.2,1.2) -- (1.2,-5.2) -- (-5.2,-5.2);
\draw[->] (-5,0) -- (1,0);
\node at (0.8,-0.3) {\small $\omega_2$};
\draw[->] (0,-5) -- (0,1);
\node at (-0.35,0.8) {\small $\gamma_2$};
\draw[blue!15!white,fill=blue!15!white] (-5,0.556) -- (0,0) -- (-5,-1.667) -- (-5,0.556);
\draw[green!15!white,fill=green!15!white] (-5,-1.667) -- (0,0) -- (-5,-5) -- (-5,-1.667);
\draw[dashed] (-5,-1.667) -- (0,0) -- (1,0.333);
\draw[dashed] (-5,0.556) -- (1,-0.111); 
\draw[ultra thick] (-5,0.556) -- (0,0); 
\draw[dashed] (-5,-5) -- (1,1);
\draw[dashed] (-5,-3.889) -- (1,0.778); 
\draw[ultra thick] (-5,-3.889) -- (0,0); 
\draw[pattern=vertical lines, pattern color=black] (-5,0.556) -- (-5,-3.889) -- (0,0) -- cycle;
\draw[pattern=horizontal lines, pattern color=black] (-5,0) -- (-5,-5) -- (0,0) -- cycle;
\draw[thick,red, ultra thick] (-5,-5) -- (0.05,0.05);
\draw[thick,red,fill=white] (0,0) circle (0.05cm);
\draw[pattern=vertical lines, pattern color=black] (-2,-3.7) -- (-1.5,-3.7) -- (-1.5,-3.3) -- (-2,-3.3) -- cycle;
\node at (-0.9,-3.5) {\scriptsize DEC};
\draw[pattern=horizontal lines, pattern color=black] (-2,-4.5) -- (-1.5,-4.5) -- (-1.5,-4.1) -- (-2,-4.1) -- cycle;
\node at (-0.9,-4.3) {\scriptsize SEC};
\node[rotate=-7] at (-4.05,0.7) {\scriptsize $\gamma_2=-\omega_2/9$};
\node[rotate=45] at (-4.1,-4.5) {\scriptsize $\gamma_2=\omega_2$};
\draw[rotate=21,black,fill=green!15!white] (-5.3,0.125) -- (-3.55,0.125) -- (-3.55,-0.18) -- (-5.3,-0.18) --(-5.3,0.125) ;
\node[rotate=19] at (-4.15,-1.6) {\scriptsize $\gamma_2=\omega_2/3$};
\draw[rotate=38,black,fill=green!15!white] (-6.35,-0.075) -- (-4.375,-0.075) -- (-4.375,-0.425) -- (-6.35,-0.425) --(-6.35,-0.075) ;
\node[rotate=37] at (-4.1,-3.505) {\scriptsize $\gamma_2=7\omega_2/9$};
\end{scope}
\end{tikzpicture}
\end{center}

\caption{Left part: Schematic overview of the conditions for $(\omega_1,\gamma_1)$ stemming from the point-wise energy conditions.  The NEC, SEC, WEC and DEC are satisfied in the coloured region, which is divided into two regions (blue and green) by the line $\gamma_1=\omega_1/2$ that indicates isotropy to leading order. The red line $\gamma_1=\omega_1$, designates the vanishing of the leading order term and conditions for subleading coefficients become important. In the right part of the Figure, the conditions for $(\omega_2,\gamma_2)$ stemming from point-wise energy conditions are visualised at $\omega_1=0=\gamma_1$.}
\label{Fig:LeadingOrder}
\end{figure}

\noindent
We next consider the FEC in eq.~(\ref{FEC}) as discussed in point \emph{(ii)} in Section~\ref{Sect:ConditionsT}. Using the series expansions, the leading order contribution implies the condition
\begin{align}
&\text{FEC:}&&\omega_1\lesssim\gamma_1<0 &&\text{or} &&0<\gamma_1\lesssim \omega_1\,.\label{FECleadingCond}
\end{align}
As before, the notation $\lesssim$ means that for $\omega_1=\gamma_1$, conditions on higher coefficients $(\omega_n,\gamma_n)$ with $n>1$ need to be satisfied for the FEC to hold.

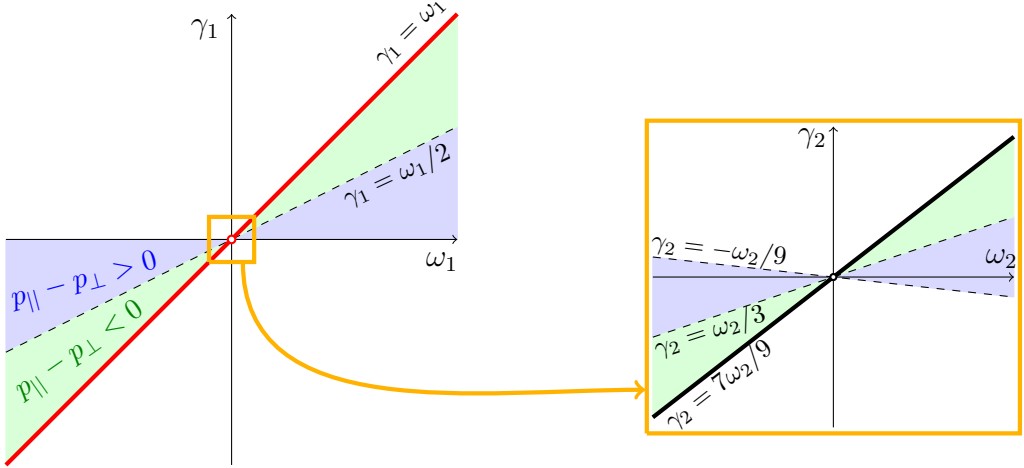
\begin{figure}[h]
\begin{center}
\begin{tikzpicture}
\draw[blue!15!white,fill=blue!15!white] (-3,0) -- (0,0) -- (-3,-1.5) -- (-3,0);
\draw[blue!15!white,fill=blue!15!white] (3,0) -- (0,0) -- (3,1.5) -- (3,0);
\draw[green!15!white,fill=green!15!white] (-3,-1.5) -- (0,0) -- (-3,-3) -- (-3,-1.5);
\draw[green!15!white,fill=green!15!white] (3,1.5) -- (0,0) -- (3,3) -- (3,1.5);
\draw[->] (-3,0) -- (3,0);
\node at (2.8,-0.3) {$\omega_1$};
\draw[->] (0,-3) -- (0,3);
\node at (-0.35,2.8) {$\gamma_1$};
\node[rotate=45] at (2.4,2.75) {\footnotesize$\gamma_1=\omega_1$};
\draw[dashed] (-3,-1.5) -- (3,1.5);
\node[rotate=25] at (2.2,0.85) {\footnotesize$\gamma_1=\omega_1/2$};
\draw[thick,red, ultra thick] (-3,-3) -- (3,3);
\draw[thick,red,fill=white] (0,0) circle (0.05cm);
\node[rotate=15,blue] at (-1.95,-0.575) {\small$p_{||}-p_\perp>0$};
\node[rotate=35,green!50!black] at (-2,-1.5) {\small$p_{||}-p_\perp<0$};
\draw[red!30!yellow,ultra thick] (-0.3,-0.3) -- (0.3,-0.3) -- (0.3,0.3) -- (-0.3,0.3) -- (-0.3,-0.3);
\draw[red!30!yellow,ultra thick,->] (0.15,-0.3) to [out=270,in=180] (5.5,-2); 
\begin{scope}[xshift=8cm,yshift=-0.5cm, scale=0.8]
\draw[red!30!yellow,ultra thick] (-3.1,-2.6) -- (3.1,-2.6) -- (3.1,2.6) -- (-3.1,2.6) -- (-3.1,-2.6);
\draw[blue!15!white,fill=blue!15!white] (-3,0.333) -- (-3,-1) -- (0,0) -- (-3,0.333);
\draw[blue!15!white,fill=blue!15!white] (3,-0.333) -- (3,1) -- (0,0) -- (3,-0.333);
\draw[green!15!white,fill=green!15!white] (-3,-1) -- (0,0) -- (-3,-2.333) -- (-3,-1);
\draw[green!15!white,fill=green!15!white] (3,1) -- (0,0) -- (3,2.333) -- (3,1);
\draw[->] (-3,0) -- (3,0);
\node at (2.8,0.3) {$\omega_2$};
\draw[->] (0,-2.5) -- (0,2.5);
\node at (-0.35,2.3) {$\gamma_2$};
\draw[dashed] (-3,0.333) -- (3,-0.333); 
\draw[ultra thick] (-3,-2.333) -- (3,2.333); 
\draw[dashed] (-3,-1) -- (3,1); 
\draw[thick,fill=white] (0,0) circle (0.05cm);
\node[rotate=38] at (-1.9,-1.8) {\footnotesize$\gamma_2=7\omega_2/9$};
\node[rotate=18] at (-2.05,-0.905) {\footnotesize$\gamma_2=\omega_2/3$};
\node[rotate=-7] at (-1.9,0.45) {\footnotesize$\gamma_2=-\omega_2/9$};
\end{scope}
\end{tikzpicture}
\end{center}

\caption{Left part: Schematic overview of the conditions for $(\omega_1,\gamma_1)$ stemming from the FEC, which is satisfied in the entire coloured region. The latter is divided into two regions (blue and green) by the line $\gamma_1=\omega_1/2$ that indicates isotropy to leading order. The red line $\gamma_1=\omega_1$, designates the vanishing of the leading order term and conditions for subleading coefficients become important. In the right part of the Figure, the conditions for $(\omega_2,\gamma_2)$ stemming from the FEC are visualised at $\omega_1=0=\gamma_1$.}
\label{Fig:LeadingOrderFEC}
\end{figure}

Graphically, the condition (\ref{LeadingDWNSEC}) is visualised in the left part of Figure~\ref{Fig:LeadingOrder} and (\ref{FECleadingCond}) in the left part of  Figure~\ref{Fig:LeadingOrderFEC}. The colouring in both Figures is related to an interpretation of the energy momentum tensor in terms of field configurations, as outlined in Section~\ref{fluid interpretation}. Indeed, $T_{\mu\nu}$ can be mimicked by a perfect fluid and either a massless scalar field or a gauge field with a non-vanishing electric component, depending on the sign of the anisotropy parameter 
 \begin{equation}
 \label{tobegiven}
     p_\parallel-p_\perp=\frac{\chi(2\gamma_1-\omega_1)}{2\pi z^4} +\mathfrak{o}(z^{-4})\ . 
 \end{equation}
 Here $\mathfrak{o}(z^{-n})$ denotes contributions more suppressed than $z^{-n}$, for example here for $n=4$ we neglect terms of the type $\log{z}/ z^5$. 
For $\omega_1=2\gamma_1$ isotropy is restored to leading order, as can be seen from \eqref{tobegiven} however, there is a residual logarithmically enhanced subleading term contributing to the anisotropy parameter as follows
 \begin{align}
&p_\parallel-p_\perp= -\frac{5\gamma_1 \chi^2 \log{z}}{8\pi z^5}+\mathcal{O}(z^{-5})>0&&\text{for} &&\gamma_1<0\,.
 \end{align}
We can therefore distinguish the two different scenarii
\begin{itemize}
\item For $p_\parallel - p_\perp < 0$  (\emph{i.e.} for $2\gamma_1 < \omega_1$)  an electric background field is needed to mimic the energy momentum tensor in eq.~(\ref{EQT}) along with an ideal fluid with the components (as parametrised in (\ref{DefIdealFluid}))
\begin{align}
\label{edispari}
&\rho=\frac{\chi(\gamma_1-\omega_1)}{2\pi z^4}+\frac{\chi^2(13\omega_1-15\gamma_1)\log{z}}{16\pi z^5}+\mathcal{O}(z^{-5})\,,  &&p= \frac{\chi^2(3\gamma_1-\omega_1)\log{z}}{16\pi z^5}+\mathcal{O}(z^{-5})\ .
\end{align}
Thus, for $\omega_1\neq \gamma_1$, $p$ is subleading in $z$ with respect to $\rho$. Therefore, asymptotically, the equation of state approaches $p=\omega \rho$ with $\omega=0$ for $\omega_1\neq \gamma_1$ and $\omega=-1$ for $\omega_1=\gamma_1$.
\item For $p_\parallel - p_\perp > 0$  (\emph{i.e.} for $2\gamma_1 \geq\omega_1$)  a scalar background field is needed to mimic the energy momentum tensor in eq.~(\ref{EQT}) along with an ideal fluid with the components (as parametrised in (\ref{DefIdealFluid}))
\begin{align}
    \label{phidispari}
 &\rho=\frac{-\gamma_1 \chi}{2\pi z^4}+\frac{3\chi^2(5\gamma_1+\omega_1)\log{z}}{16\pi z^5}+\mathcal{O}(z^{-5}),&& p= \frac{\chi^2(3\gamma_1-\omega_1)\log{z}}{16\pi z^5}+\mathcal{O}(z^{-5})\  .
\end{align}
As for the previous case, asymptotically, the equation of state approaches $p=\omega \rho$ with $\omega=0$ for $\omega_1\leq \gamma_1<0$.

\end{itemize}

\subsubsection{Case $(\omega_1,\gamma_1)=(0,0)$ and Higher Orders}\label{EC2}
In the previous subsection we have excluded the case $(\omega_1,\gamma_1)=(0,0)$, \emph{i.e.} the origin in the left part of Figures~\ref{Fig:LeadingOrder} and \ref{Fig:LeadingOrderFEC}. However, it is instructive to investigate the limit in which the leading corrections to $f$ and $h$  emerge at the order $1/d^2$. This is the first relevant order in some quantum extensions of Black Hole metrics that have been considered in the literature \cite{Bonanno:2000ep,Binetti:2022xdi}. Indeed, the restriction $\omega_1=\gamma_1=0$ can be motivated from effective actions of quantum gravity, whose leading terms appear at the two-derivative level \cite{Donoghue:1994dn,Donoghue:1995cz,Donoghue:2022chi,Bjerrum-Bohr:2002gqz}. In the particular case $(\omega_2,\gamma_2)\neq (0,0)$, the eigenvalues of the Einstein tensor to the first non-vanishing order now read 
\begin{align}
 &\epsilon=-\frac{\chi  \omega_2 }{2 \pi  z^5}+\mathfrak{o}(z^{-5})\,,&&p_\parallel=\frac{\chi  (3 \gamma_2 -\omega_2 )}{4 \pi  z^5}+\mathfrak{o}(z^{-5})\ ,&&p_\perp&\approx \frac{3 \chi  (\omega_2 -3 \gamma_2 )}{8 \pi  z^5}+\mathfrak{o}(z^{-5})\ \label{pper}.
\end{align}
Inserting these expansions into the point-wise energy conditions lead to the following conditions
\begin{align}
&\text{DEC:}\hspace{0.5cm}\frac{7}{9}\,\omega_2<\gamma_2<-\frac{\omega_2}{9}\,,&&\text{WEC:} \hspace{0.5cm}\omega_2\lesssim\gamma_2<-\frac{\omega_2}{9}\,,\nonumber\\
&\text{NEC:} \hspace{0.5cm}\omega_2\lesssim\gamma_2<-\frac{\omega_2}{9}\,,&&\text{SEC:} \hspace{0.7cm}\omega_2\lesssim\gamma_2\lesssim 0\,.\label{CondsDWNS}
\end{align}
These conditions are schematically visualised in the right part of Figure~\ref{Fig:LeadingOrder}. The FEC condition in this case takes the form
\begin{align}
&\text{FEC:} &&\frac{7\omega_2}{9}< \gamma_2\leq-\frac{\omega_2}{9}\,,&&\text{or} &&-\frac{\omega_2}{9}\leq\gamma_2< \frac{7\omega_2}{9}\,,
\end{align}
which is visualised in the right hand side of Figure~\ref{Fig:LeadingOrderFEC}. In Figures~\ref{Fig:LeadingOrder} and \ref{Fig:LeadingOrderFEC} we have also indicated the sign of the anisotropy parameter $p_{||}-p_\perp$ in this case, which is positive for $\gamma_2<\omega_2/3$ and negative else. In this case, with the constraints given by the NEC, WEC, SEC, the stress-energy tensor can only be mimicked by an electric background  field $(\omega_2\leq \gamma_2<\frac{\omega_2}{3})$ and a perfect fluid component given by an equation of state of the form $p=\omega \rho $, with $\omega=-1$. For the remaining conditions (DEC, FEC), for non-zero $\gamma_2,\omega_2$, the realisation of the stress-energy tensor in terms of an isotropic fluid plus a perfect fluid contribution is unavailable.

More generally, for $(\omega_{i},\gamma_i)=(0,0)$ for $i=1,\ldots,n-1$ and $(\omega_n,\gamma_n)\neq (0,0)$, we have verified up to $n=5$ that the point-wise energy conditions imply\footnote{For $n>2$ the NEC, WEC and DEC even hold if the upper bound for $\gamma_n$ is satisfied.}
\begin{align}
&\text{DEC:}\hspace{0.5cm}\frac{3n+1}{(n+1)^2}\,\omega_n<\gamma_n<-\frac{n-1}{(n+1)^2}\,\omega_n\,,&&\text{WEC:}\hspace{0.5cm}\omega_n\lesssim\gamma_n<-\frac{n-1}{(n+1)^2}\,\omega_n\,,\nonumber\\
&\text{NEC:}\hspace{0.5cm}\omega_n\lesssim\gamma_n<-\frac{n-1}{(n+1)^2}\,\omega_n\,,&&\text{SEC:}\hspace{0.7cm}\omega_n\lesssim\gamma_n\lesssim0\,,\label{WNSDECsub}
\end{align}
while the FEC generalises to 
\begin{align}
&\text{FEC:} &&\frac{3n+1}{(n+1)^2}\omega_n\leq \gamma_n<-\frac{n-1}{(n+1)^2}\omega_n\,,&&\text{or} &&-\frac{n-1}{(n+1)^2}\omega_n<\gamma_n\leq \frac{3n+1}{(n+1)^2}\omega_n\,.\label{FECsub}
\end{align}

\subsubsection{Case $f=h$}
\label{f=h}
We have seen in Section~\ref{Sect:EnergyConditionsGen} that the point-wise energy conditions lead to much simpler results for the particular case $f=h$. Similarly, we have seen in the previous Subsubsection~\ref{EC2} that the case $\omega_n=\gamma_n$ for the first non-trivial coefficients requires to verify subleading terms in order to determine whether most of the energy conditions are satisfied or not. In this Subsubsection, we shall therefore consider in more detail the particular case $\omega_n=\gamma_n$ $\forall n \in\mathbb{N}$. 

For $\omega_i=0$ $\forall i=1,\ldots,n-1$ and $\omega_{n}\neq 0$, we have verified up to $n=5$ that the eigenvalues of the energy momentum tensor can be expanded as
\begin{align}
&\epsilon=-p_{||}=-\frac{\chi\, n\, \omega_n}{4\pi z^{3+n}}+\mathfrak{o}(z^{-n-3})\,,&&\text{and} &&p_\perp=-\frac{\chi\, n\,(n+1)\,\omega_n }{8\pi z^{3+n}}+\mathfrak{o}(z^{-3-n})\,.
\end{align}
In this case, we can therefore conclude that the NEC, WEC, DEC and SEC are satisfied provided that $\omega_n< 0$, \emph{i.e.} the first non-vanishing coefficient has to be negative. Likewise, the FEC is satisfied only if $\omega_1\neq 0$, in agreement with the general discussion of Section~\ref{Sect:RelFEC}: indeed for $\omega_1\neq 0$ we have the expansion
\begin{align}
\phi(z)=1+\frac{\omega_1}{z}-\frac{\chi\,\omega_1}{z^2}\,\log(z)+\mathcal{O}(z^{-2})\,,
\end{align}
such that the sign of the first correction beyond the term of order $\mathcal{O}(z^{-1})$ is opposite of the latter.


\subsection{Energy Conditions at Finite Distance}
\label{logaritmo}
Extending the analysis of the previous Subsection to finite distance (\emph{i.e.} to higher orders in an expansion of $z$) is a challenging task: not only do we require an expansion of the distance $d$ as a function of $z$ (see eq.~(\ref{DefSeriesExpansion})) to higher orders, but we also need to characterise the region in which such an approximation is valid. In fact, as we shall see, both aspects are intertwined. In this work we shall not attempt an exhaustive discussion of all (point-wise) energy conditions for finite (but large) distance. Instead, we shall simply illustrate the two points mentioned above by two examples.

\subsubsection{Example: NEC at Higher Orders}\label{Sect:CordConstraints}
In order to showcase the idea (and also to highlight the problems) to extend the point-wise energy conditions beyond the leading asymptotic distance, we shall first consider as a simple example the NEC condition in the case of a metric with $f=h$. We furthermore consider a critical distance $d_c > d_A$ (see Figure~\ref{Fig:ScaleBH}), such that the series expansions (\ref{Functionsfh}) of the metric functions are not valid all the way to $\za$ but only hold in the asymptotic regime. That is, the red region in Figure~\ref{Fig:ScaleBH} does not extend all the way to $\za$. In the following we shall attempt to develop conditions on the expansion coefficients $\omega_n=\gamma_n$ by demanding that the NEC is satisfied in this regime.

\begin{figure}[htbp]
\begin{center}
\includegraphics[width=8cm]{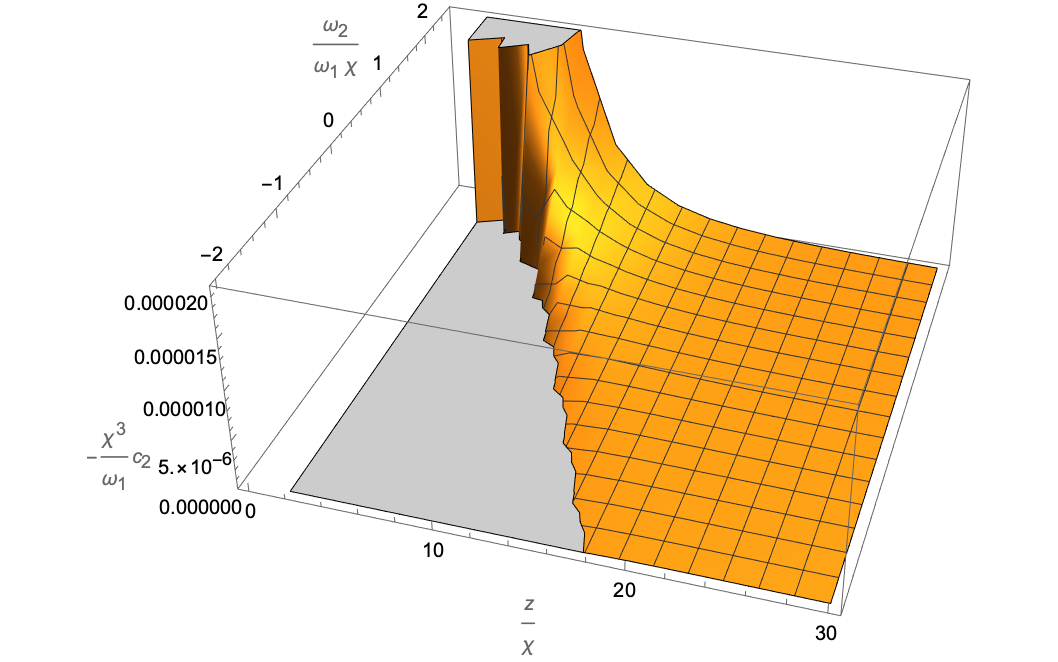}\hspace{0.5cm} \includegraphics[width=8cm]{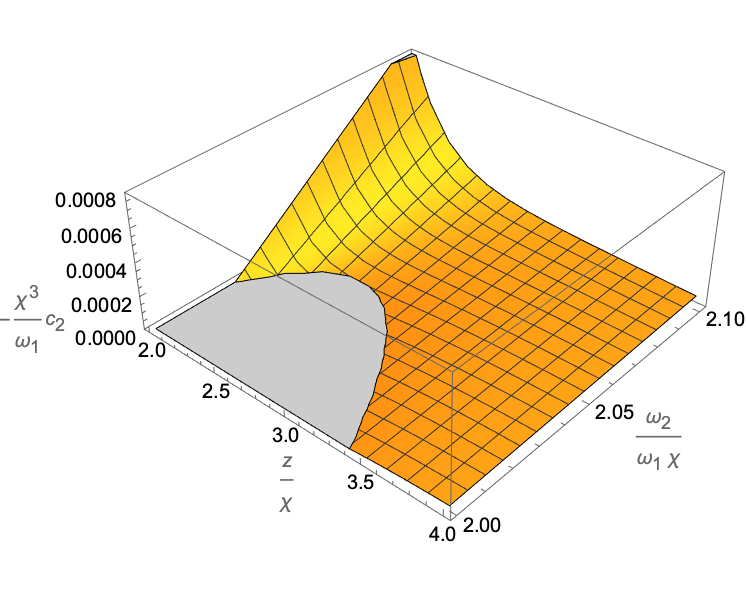}
\end{center}
\caption{\sl Left panel: condition $-\frac{\chi^3}{\omega_1}\,\mathfrak{c}_2$ as a function of $\frac{z}{\chi}$ and $\frac{\omega_2}{\omega_1\chi}$. For negative values of $\omega_1$ (as dictated from the asymptotic constraint (\ref{LeadingDWNSEC})), the orange coloured part of the plot indicates the parameter region in which the NEC is satisfied. Right panel: More detailed view of the same plot, highlighting the region of the zero of $\mathfrak{c}_2$.}
\label{Fig:NoScaleCondition}
\end{figure}

\begin{figure}[htbp]
\begin{center}
\includegraphics[width=9cm]{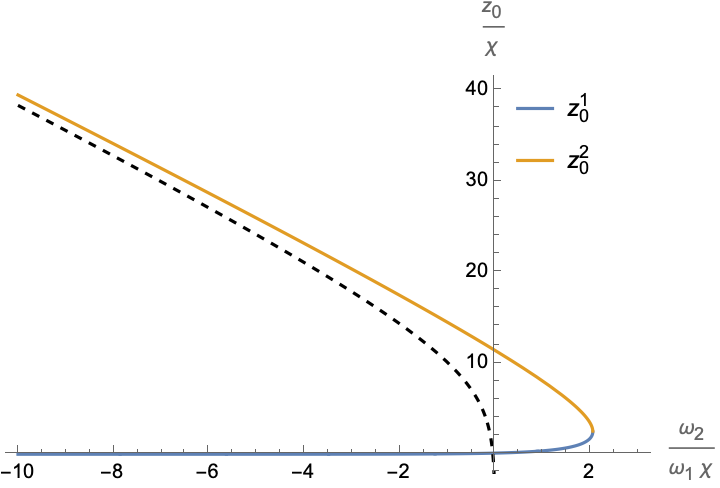}
\end{center}
\caption{\sl Two branches of the position of the zero $z_0/\chi$ of $\mathfrak{c}_2$ (to order $\mathcal{O}(z^{-6})$) as a function of $\frac{\omega_2}{\chi\omega_1}$ as in eq.~\ref{PosZero}, which meet at $\frac{\omega_2}{\chi\omega_1}=-\frac{27}{10}+\pi+\log5$ for $z_0=\frac{5}{2}\,\chi$. The black dashed lines shows the asymptotic curve given in (\ref{AsymPos}).}
\label{Fig:NoScaleZero}
\end{figure}

In the case of $f=h$, the first condition in (\ref{ec1}) $\mathfrak{c}_1=0$ is identically satisfied (and poses no conditions on the coefficients $\omega_n$). Assuming furthermore $\omega_1\neq 0$ the series expansion of $\mathfrak{c}_2$ in (\ref{ec1}) becomes\footnote{The terms of order $\mathcal{O}(z^{-6})$ can be logarithmically enhanced, which, however, does not interfer with our analysis below.}
\begin{align}
\mathfrak{c}_2=-\frac{\chi\omega_1}{2\pi z^4}+\frac{\chi^2\omega_1}{8\pi z^5}\left[10\pi-17+10\log\left(\frac{2z}{\chi}\right)-\frac{10\omega_2}{\chi\omega_1}\right]+\mathcal{O}(z^{-6})\,,\label{ExpandCondc2}
\end{align}
which is plotted in Figure~\ref{Fig:NoScaleCondition} as a function of $z/\chi$ and $\frac{\omega_2}{\omega_1\chi}$. As a function of $z$, $\mathfrak{c}_2$ has zeros at
\begin{align}
z_0=-\frac{5}{2}\,\chi\,W\left(-\frac{1}{5}\,e^{\frac{17}{10}-\pi+\frac{\omega_2}{\chi\omega_1}}\right)\,,\label{PosZero}
\end{align}
where $W$ is the Lambert function: for 
\begin{align}
\frac{\omega_2}{\chi\omega_1}\in\bigg]-\infty,-\frac{27}{10}+\pi+\log5\bigg]\,,
\end{align}
eq.~(\ref{PosZero}) in fact describes two real solutions (corresponding to the two branches of the Lambert function), which meet at $z_0=\frac{5}{2}$, while for $\frac{\omega_2}{\chi\omega_1}>-\frac{27}{10}+\pi+\log5$, the solution $z_0$ is imaginary. The position of the real zeroes are plotted in Figure~\ref{Fig:NoScaleZero} as a function of $\frac{\omega_2}{\chi\omega_1}$. For large negative $\frac{\omega_2}{\omega_1\chi}$, the larger of the two zeroes behaves as
\begin{align}
-\frac{5}{2}\,\frac{\omega_2}{\omega_1\chi}+\frac{1}{4}\left(10\pi-17+10\log\left(-\frac{5\omega_2}{\omega_1\chi}\right)\right)+\mathcal{O}\left(-\frac{\omega_1\chi}{\omega_2}\right)\,.\label{AsymPos}
\end{align}

\noindent
There are a number of conclusions to be drawn from this simple analysis:
\begin{itemize}
\item as outlined in Section~\ref{NECLeadingOrder}, the leading asymptotic expansion (\emph{i.e.} the term of order $\mathcal{O}(z^{-4})$) requires $\omega_1<0$ (since we assumed $\omega_1\neq 0$)
\item for $\omega_2>\left(-\frac{27}{10}+\pi+\log 5\right)\chi\omega_1$ (and $\omega_1<0$), $\mathfrak{c}_2\geq 0$ (and thus the NEC) is satisfied up to order $\mathcal{O}(z^{-6})$
\item for $\omega_2\leq \left(-\frac{27}{10}+\pi+\log 5\right)\chi\omega_1$ (and $\omega_1<0$), the NEC is violated for values of $z$ below $z_0$ given in (\ref{PosZero}). By tuning $\omega_2$, this value can in particular be made larger than $z_c$ (and also large enough such that the restriction to order $\mathcal{O}(z^{-6})$ in (\ref{ExpandCondc2}) is justified). This therefore imposes a non-trivial restriction for $\omega_2$
\end{itemize}
This analysis, however, also reveals a number important subtleties: 
\begin{enumerate}
\item[{\emph{(i)}}] The restriction for $\omega_2$ involves $z_c$ as a value up to which we  expect the expansions of the metric function to be still a viable approximation. This, however, is a coordinate-dependent statement which is difficult to formulate as a physical condition. 
\item[{\emph{(ii)}}] In order to violate the NEC, the quotient $\omega_2/\omega_1$ needs to scale proportional to $\chi$: \emph{e.g.} for an astrophysical black hole $\chi\gg 1$, such that the $\omega_n$ cannot be treated as coefficients of order unity. 
\item[{\emph{(iii)}}] For the expansion (\ref{Functionsfh}) to have radius of convergence $u_c$, the $\omega_n$ themselves have to scale relative to one-another by factors that are comparable to $u_c$, which we need to take into account when performing expansions.  This case will be considered later in Section~\ref{Sect:RadiusConvergence}.
\end{enumerate}

\noindent
For completeness, we remark that the restriction $f=h$ can be relaxed in a straight-forward manner: In this case, $\mathfrak{c}_1$ is no longer trivial and the two conditions in (\ref{ec1}) become\footnote{As before, the subleading terms of order $\mathcal{O}(z^{-6})$ may be enhanced logarithmically. }
\begin{align}
0\leq \mathfrak{c}_1&=\frac{\chi (\gamma_1-\omega_1)}{2\pi z^4}-\frac{3\chi}{4\pi z^5}\left[(\gamma_1-\omega_1)\chi(\pi-4/3)+\omega_2-\gamma_2+\chi(\gamma_1-\omega_1)\,\log\left(\frac{2z}{\chi}\right)\right]+\mathcal{O}(z^{-6})\,,\label{SubleadingC1}\\
0\leq \mathfrak{c}_2&=-\frac{\gamma_1\chi }{2\pi z^4}+\frac{\chi}{8\pi z^5}\left[\chi(9\gamma_1+\omega_1)\left(\pi+\log\left(\frac{2z}{\chi}\right)-(15\gamma_1+2\omega_1)-(9\gamma_2+\omega_2)\right)\right]+\mathcal{O}(z^{-6})\,.\label{SubleadingC2}
\end{align}
We highlight again that these constraints heavily depend on the choice of coordinates along with the point $z_c$ up to which we trust the approximation. Indeed, while the asymptotically leading terms (already discussed in Section~\ref{NECLeadingOrder}), require $\omega_1\lesssim\gamma_1<0$ (see eq.~(\ref{LeadingDWNSEC})), neglecting the higher order terms $\mathcal{O}(z^{-6})$, the functions in (\ref{SubleadingC1}) and (\ref{SubleadingC2}) have zeroes at
\begin{align}
&\mathfrak{c}_1:&&z_0^{(1)}=-\frac{3}{2}\,\chi\,W\left[-\frac{1}{3}\text{exp}\left(\frac{3(\gamma_2-\omega_2)+(4-3\pi)(\gamma_1-\omega_1)\chi}{3\chi(\gamma_1-\omega_1)}\right)\right]\,,\nonumber\\
&\mathfrak{c}_2: &&z_0^{(2)}=-\frac{9\gamma_1+\omega_1}{4\gamma_1}\,\chi\,W\left[-\frac{3\gamma_1}{9\gamma_1+\omega_1}\,\text{exp}\left(\frac{\chi\left((15-9\pi)\gamma_1+(2-\pi)\omega_1\right)+9\gamma_2+\omega_2}{\chi(9\gamma_1+\omega_1)}\right)\right]\,.\nonumber
\end{align}
A better intuition for these zeroes can be obtained by expanding them for large $\chi$\footnote{In the case of an astrophysical black hole, $\chi\gg 1$, such that such an expansion is well justified.}, assuming that $(\omega_n,\gamma_n)$ are independent of $\chi$
\begin{align}
&z_0^{(1)}=\frac{3\lambda_1\chi}{2}+\frac{3(\gamma_2-\omega_2)}{2(1-\lambda_1)(\gamma_1-\omega_1)}+\frac{3\lambda_1(\gamma_2-\omega_2)^2}{4(1-\lambda_1)^3(\gamma_1-\omega_1)^2\chi}+\mathcal{O}(\chi^{-2})\,,\nonumber\\
&z_0^{(2)}=\frac{(9\gamma_1+\omega_1)\lambda_2\chi}{4\gamma_1}+\frac{(9\gamma_2+\omega_2)\lambda_2}{4(1-\lambda_2)\gamma_1}+\frac{(9\gamma_2+\omega_2)^2\lambda_2}{8\gamma_1(1-\lambda_2)^3(9\gamma_1+\omega_1)\chi}+\mathcal{O}(\chi^{-2})\,,
\end{align}
with the coefficients
\begin{align}
 &\lambda_1=-W\left(-\frac{e^{\frac{4}{3}-\pi}}{3}\right)\sim0.058\,,&&\lambda_2=-W\left(-\frac{2\gamma_1}{9\gamma_1+\omega_1}\text{exp}\left(\frac{15\gamma_1+2\omega_1-\pi(9\gamma_1+\omega_1)}{9\gamma_1+\omega_1}\right)\right)\,.\nonumber
\end{align}
As before, the zeroes $z_0^{(1,2)}$ depend on $(\gamma_n,\omega_n)$ and can in principle be made larger than $z_c$, leading to non-trivial (albeit implicit) conditions.


\subsubsection{Example: Radius of Convergence}\label{Sect:RadiusConvergence}
One possibility of fixing the distance $d_c$ in Figure~\ref{Fig:ScaleBH} is to identify it with the radius of convergence of the series appearing in (\ref{Functionsfh}). For concreteness, we consider again the simpler case $f(z)=h(z)$ $\forall z>z_c$  (\emph{i.e.} $\omega_n=\gamma_n$ $\forall n\in\mathbb{N}$) and we write
\begin{align}
&f(z)=h(z)=1-\frac{2\chi}{z}\left( 1 + \sum_{n=1}^\infty \omega_n\,u^n(z) \right)=1-\frac{2\chi}{z}\left(1+\sum_{n=1}^\infty\alpha_n\,d_c^n\,u^n\right)\,,&&\text{with} &&u(z)=\frac{1}{d(z)}\,.\label{RelSeries}
\end{align} 
Here we have also re-scaled the expansion coefficients
\begin{align}
&\alpha_n:=\omega_n\,u_c^n=\omega_n/d_c^n\,,&&\text{with} &&u_c=1/d_c\,,
\end{align}
where $\{\alpha_n\}_{n\in\mathbb{N}}$ are the coefficients of a series with radius of convergence equal to 1. In this way, the radius of convergence of the coefficients $\omega_n$ indeed is (assuming that these limits exist)
\begin{align}
\lim_{n\to\infty}\left|\frac{\omega_n}{\omega_{n+1}}\right|=\lim_{n\to\infty}\left|\frac{\alpha_n/u_c^n}{\alpha_{n+1}/u_c^{n+1}}\right|=u_c\lim_{n\to\infty}\left|\frac{\alpha_n}{\alpha_{n+1}}\right|=u_c\,.
\end{align}
Therefore $1/u_c$ is not only the smallest distance up to which we assume that the series expansion (\ref{RelSeries}) converges against the actual metric functions $f$, but also up to which these series are in fact meaningful. Assuming $u_c$ to be small (such that $d_c$ is large), we can study the NEC conditions (\ref{ec1}). Condition $\mathfrak{c}_1$ is identically satisfied (since we have assumed $f=h$) and it therefore remains to study the condition under which $\mathfrak{c}_2$ is positive when evaluated at $d_c$.\footnote{We assume that eq.~(\ref{LeadingDWNSEC}) (respectively eq.(\ref{WNSDECsub})) hold, such that the NEC is satisfied asymptotically.} To this end, we insert the leading terms in $z$ in (\ref{Expansionzd}) into the contribution proportional to $\omega_n$ of the $\mathcal{O}(z^{-3-n})$ term in an expansion of the condition $\mathfrak{c}_2$
\begin{align}
\mathfrak{c}_2=-\frac{\chi}{8\pi}\,u_c^3\,\sum_{n=1}^\infty n(n+3)\,\alpha_n+\mathcal{O}(u_c^4)\,.\label{Relc21}
\end{align}  
This relation receives corrections from higher order terms, which, however, are more difficult to compute. Nevertheless, we have checked the following relation up to $n=5$
\begin{align}
\mathfrak{c}_2=-\frac{\chi}{8\pi}\,u_c^3\,\sum_{n=1}^\infty n(n+3)\,\alpha_n-\frac{\chi^2}{8\pi}\,u_c^4\sum_{n=1}^\infty\left[n^2+\left(\pi+\log\left(\frac{2}{\chi u_c}\right)n(n+5)\right)\right]\alpha_n+\mathcal{O}(u_c^5)\,.\label{Relc22}
\end{align}
Yet higher terms in an expansion of $u_c$ depend on the $\alpha_n$ in a non-linear fashion. Positivity of the relation (\ref{Relc21}) (or higher order terms of the form (\ref{Relc22})) pose conditions on the coefficients $\alpha_n$ stemming from the fact that the series expansion (\ref{RelSeries}) has to hold up to finite distance $u_c=1/d_c$ to the origin.

\section{Examples}\label{Sect:Examples}
In this Section we compare our energy conditions to examples in the literature that were proposed to describe quantum black holes. Specifically, we focus on the Dymnikova space-time \cite{dymnikova1992vacuum} and the asymptotically safe quantum black hole \cite{Bonanno:2000ep}.
\subsection{The Dymnikova  Black Hole} 
The Dymnikova \cite{dymnikova1992vacuum} space-time provides an explicit form for the function $f(z)=h(z)$ in \eqref{metrica adimensionale} (and thus also for $\phi(z)=\psi(z)$). It was constructed as a black hole solution of the Einstein equation \eqref{EQT} that resolves the singularity at the origin. Indeed, the dimensionless metric \cite{dymnikova1992vacuum} reads
\begin{align}
\label{metricadn}
    \dd \sigma^2&=-\left(1-\frac{R_S(z)}{z}\right)\dd \tau^2+\frac{\dd z^2}{1-\frac{R_S(z)}{z}}+z^2\left(\dd\theta^2+\sin^2\theta\dd\phi^2\right)\\
    R_S(z)&=z_S\left(1-e^{-\frac{z^3}{z^3_\ast}}\right)\qq{with} z^3_\ast=z^2_0 z_S \ ,\qq{and} z_S=2\chi\,,
\end{align}
such that we have in comparison to (\ref{MetricFctGen})
\begin{equation} 
\phi(z) =  \psi(z)=\frac{R_S(z)}{z_S}=1-e^{-\frac{z^3}{z_*^3}}\leq 1\ .\label{FormPhiDymnikova}
\end{equation} 
The metric (\ref{metricadn}) has two horizons (\emph{i.e.} the function $1-\frac{R_S(z)}{z}$ has two zeroes) at
\begin{align}
&z_+=2\chi\left[1-\mathcal{O}\left(e^{-\frac{4\chi^2}{z_0^2}}\right)\right]<2\chi \ ,&&\text{and} &&z_-=2\chi\left[1-\mathcal{O}\left(\frac{z_0}{8\chi}\right)\right]\,,
\end{align} 
where $z_H=z_+$ is the outer horizon of the black hole. Asymptotically (\emph{i.e.} for $z/z_*\rightarrow \infty$), the

\begin{wrapfigure}{l}{0.40\textwidth}
${}$\\[-1cm]
\begin{center}
\includegraphics[width=6.5cm]{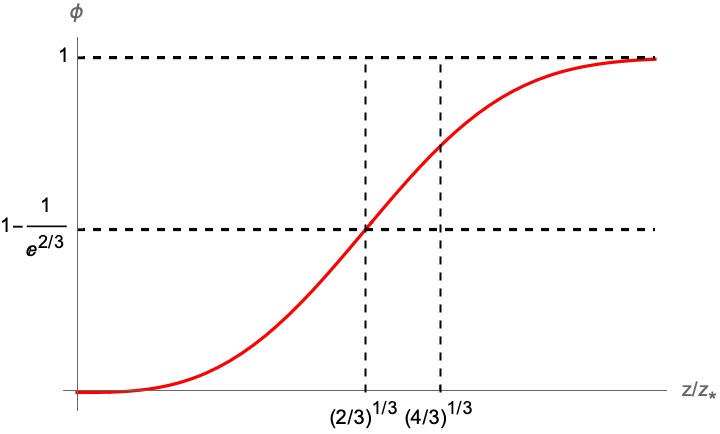}
\end{center}
\caption{Schematic plot of the function $R_S(z)/z_S$ in eq.~(\ref{FormPhiDymnikova}). }
\label{Fig:SchemFuncPhi}
${}$\\[-1cm]
\end{wrapfigure}

\noindent
metric approaches flat space (like the classical Schwarzschild solution), but it is free from a singularity at the origin $z=0$. Finally, we identify the constant $z_0^2=\frac{3}{8\pi \epsilon_0}$, where $\epsilon_0>0$ is the energy density evaluated at the origin $z=0$.

The energy momentum tensor (\ref{EQT}) associated with the space-time (\ref{metricadn}) is spherically symmetric and anisotropic \cite{dymnikova1992vacuum} and of type I in the Ellis-Hawking classification \cite{Hawking:1973uf} (see also the review in appendix~\ref{App:EllisHawking}). Using the decomposition (\ref{emtensor}) in terms of the eigenvalues  (\ref{FormT}) it can be characterised by
\begin{align}
&\epsilon= \epsilon_0\, e^{-\frac{4\pi z^3 \epsilon_0}{3\chi}}= -p_\parallel\,,&&p_\perp=-\epsilon_0 e^{-\frac{4\pi z^3 \epsilon_0}{3\chi}} \left(1-\frac{2\pi z^3 \epsilon_0}{\chi}\right) \ .
\end{align} 
These quantities allow us to verify the various energy conditions discussed in Section~\ref{Sect:EnergyConditionsGen}. Since $\epsilon\geq 0$ and $\epsilon\geq p_\perp$, the NEC and WEC are satisfied in the entire space-time. The DEC and SEC, however, require the following conditions
\begin{align}
&\text{SEC:}\hspace{0.15cm}\frac{4\pi\epsilon_0^2}{\chi}\,e^{-\frac{4\pi\epsilon_0 z^3}{3\chi}}\,\left(z^3-\frac{2}{3}\,z_*^3\right)\geq 0\,,&&\text{DEC:}\hspace{0.15cm}\frac{2\pi\epsilon_0^2}{\chi}\,e^{-\frac{4\pi \epsilon_0}{3\chi}}\,\left(\frac{2}{3}\,z_*^3-\left|z^3-\frac{2}{3}\,z_*^3\right|\right)\geq 0\,,
\end{align}
while the FEC amounts to
\begin{align}
\text{FEC:}\hspace{0.15cm}-\frac{4\pi^2 z^3\epsilon_0^4}{\chi^2}\,e^{-\frac{8\pi\epsilon_0 z^3}{3\chi}}\,\left(z^3-\frac{4}{3}\,z_*^3\right)\geq 0\,.
\end{align}
The SEC is therefore only satisfied for $z\geq  (\tfrac{2}{3})^{1/3}\,z_*$. The second derivative of the function $\phi$ vanishes at $z=(\frac{2}{3})^{1/3}\,z_*$, (\emph{i.e.} $\phi''(z=(\frac{2}{3})^{1/3}\,z_*)=0$) and it is negative for all values of $z$ larger than this. This is therefore compatible with the relation (\ref{ineqPhi}), which we have found on general grounds and the schematic drawing of the function $\phi$ in Figure~\ref{Fig:SchemFuncPhi} for $z>(\frac{2}{3})^{1/3}\,z_*$ is indeed compatible with the general form of $\phi$ we have sketched in Figure~\ref{Fig:SchemFuncPhiSEC}. 

The DEC is only satisfied for $\left(\frac{4}{3}\right)^{1/3}\, z_*\geq z\geq \left(\frac{2}{3}\right)^{1/3}\,z_*$ and is notably violated asymptotically for large $z$. Similarly, the FEC only holds for $z\leq \left(\frac{4}{3}\right)^{1/3}\,z_*$, but not for asymptotically large values of $z$. This is compatible with the general conclusion in Section~\ref{Sect:RelFEC}:  the function $\phi(z)-1$ in (\ref{FormPhiDymnikova}) tends to zero exponentially (and thus faster than $1/z$). The norm of the energy flow, however, remains bounded from below: indeed, the quantity $\epsilon^2-p_\perp^2$ has a minimum at $z=2^{1/3}\,z_*$ such that
\begin{align}
&\epsilon^2-p_{||}^2=0\,,&&\text{and} &&\epsilon^2-p_\perp^2\geq -\frac{27\chi^2}{16 e^4 \pi^2\,z_*^6}=-\frac{3\epsilon_0^2}{e^4}\,,&&\forall z\geq 0\,.
\end{align}
We also remark that the violation of both the DEC and FEC decreases exponentially for $z\gg 1$.

The anisotropy parameter, which characterises how to interpret the energy momentum tensor in eq.~(\ref{EQT}) in terms of simple physical field configurations (see Section~\ref{fluid interpretation}) becomes 
\begin{equation}
   p_\parallel-p_\perp= -\frac{2 \pi  z^3 \epsilon_0^2 e^{-\frac{4 \pi  z^3 \epsilon_0}{3 \chi }}}{\chi } < 0\ .
\end{equation}
In agreement with (\ref{Anisotropy}), this quantity is always negative and therefore, in terms of the minimal particle physics realisation, the Einstein tensor associated with the metric (\ref{metricadn}) can be mimicked by an electric-like background field whose modulus is given by \eqref{elec}
\begin{equation}
    \abs{E}=\sqrt{\frac{2 \pi  z^3 \epsilon_0^2 e^{-\frac{4 \pi  z^3 \epsilon_0}{3 \chi }}}{\chi }}\ .
\end{equation}
The energy density and pressure of the isotropic component obtained using \eqref{rho1} yield
\begin{align}
    &\rho =\frac{\epsilon_0 e^{-\frac{4 \pi  z^3 \epsilon_0}{3 \chi }} \left(\chi -\pi  z^3 \epsilon_0\right)}{\chi }=-p\,,
\end{align}
in agreement with the conclusion (\ref{EoSGenfh}) we have obtained on more general grounds. Thus, according to Table \ref{fluido1}, the anisotropic fluid associated with the Dymnikova space-time is minimally modelled by an electric-like background and massive scalar field with negligible kinetic term at large distances.

\subsection{Renormalisation Group Improved Black Hole}
Another example for a quantum deformed black hole solution, which is closer in spirit to the form (\ref{Functionsfh}) of the metric functions, is the renormalisation group (RG)-improved black hole discussed in \cite{Bonanno:2000ep}. Concretely, the latter is characterised by the following metric functions in eq.~(\ref{metrica adimensionale})
\begin{align}
&f(z)=h(z)=1-\frac{2\chi}{z}\,\frac{\widetilde{d}(z)^2}{\widetilde{d}(z)^2+\widetilde{\omega}}\,,
&&\text{with} &&\widetilde{d}(z)=\int_0^z\frac{dz'}{\sqrt{|f(z')|}}\,.\label{RGASdef}
\end{align}
Here $\widetilde{\omega}\neq 0$ is a dimensionless constant (which in \cite{Bonanno:2000ep} was determined to be $\widetilde{\omega}=\frac{118}{15\pi}$). We also note that the physical distance $\widetilde{d}(z)$ is different than the one appearing in (\ref{DefDistanceAsymp}). Indeed, since (\ref{RGASdef}) holds in the entire space-time (\emph{i.e.} $\forall z\geq 0$), the distance is normalised such that $\widetilde{d}(z=0)=0$. As we have briefly explained in Appendix~\ref{App:SchemeDependence}, this difference has no conceptual significance.

For large enough distances to the center of the black hole, such that $\widetilde{\omega}\leq \widetilde{d}(z)^2$, the (exact) metric function (\ref{RGASdef}) can be expanded in a series similar to (\ref{RelSeries})
\begin{align}
&f(z)=1-\frac{2\chi}{z}\,\sum_{n=0}^\infty\frac{(-\widetilde{\omega})^n}{\widetilde{d}(z)^{2n}}=1-\frac{2\chi}{z}\,\sum_{n=0}^\infty(-\,\text{sign}(\widetilde{\omega}))^{n/2}|\widetilde{\omega}|^{n/2}\, \widetilde{u}^n\,,&&\text{with} &&\widetilde{u}(z)=\frac{1}{\widetilde{d}(z)}\,.\label{ASseries}
\end{align}
Furthermore, asymptotically, the distance behaves (see \cite{Bonanno:2000ep}) as $\widetilde{d}(z)\sim z$, such that 
\begin{align}
&\epsilon=-p_\parallel= \frac{\chi  \tilde{\omega} }{2 \pi  z^5}+\mathcal{O}(z^{-6})\,,&&p_\perp= \frac{3 \chi  \tilde{\omega}}{4 \pi  z^5}+\mathcal{O}(z^{-6})\,.
\end{align}
Asymptotically, for $\widetilde{\omega}>0$, on the one hand the NEC, WEC and SEC are satisfied in agreement with eq.~(\ref{CondsDWNS}), while on the other hand the DEC and FEC are violated in agreement with eq.~(\ref{CondsDWNS})) and the discussion in Section~\ref{Sect:RelFEC}: indeed, the derivative of $\frac{\widetilde{d}(z)^2}{\widetilde{d}(z)^2+\widetilde{\omega}}$ tends to zero faster than $\mathfrak{o}(z^{-2})$ for $z\to \infty$. Finally, asymptotically, the anisotropy parameter becomes
\begin{equation}
   p_\parallel-p_\perp=-\frac{5\chi \tilde{\omega}}{4 \pi z^5}<0\ ,
\end{equation}
in agreement with (\ref{Anisotropy}). For large distances, the energy momentum associated with (\ref{RGASdef}) can therefore mimicked by an electric field and a perfect fluid with equation of state $p=-\rho$, as shown in eq.~(\ref{EoSGenfh}).

\begin{figure}[htbp]
\begin{center}
\includegraphics[width=7.5cm]{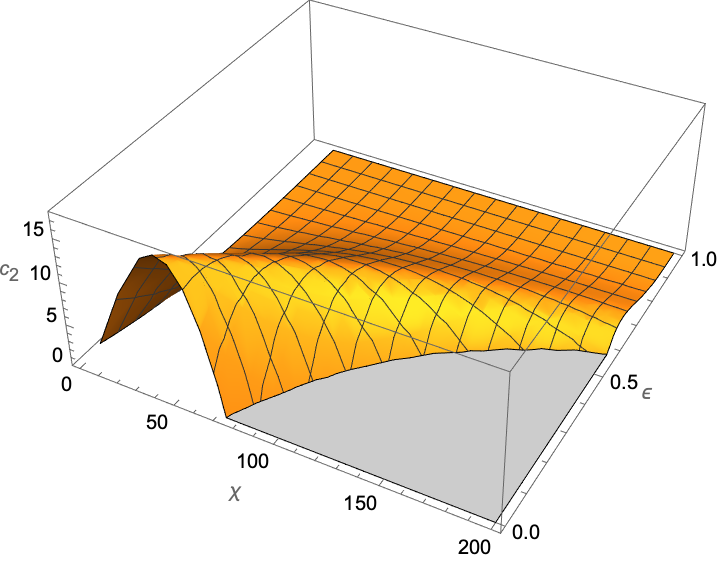}
\end{center}
\caption{\sl The condition $\mathfrak{c}_2$ as a function of $\chi$ and $\epsilon$ for $\widetilde{\omega}=\frac{118}{15\pi}$. The orange coloured region gives the regions where $\mathfrak{c}_2$ is positive at $u_c$. Notice that for $\epsilon\ll 1$, higher order terms in (\ref{Relc22}) cannot be neglected.}
\label{Fig:ConditionRelRadiusConvergence}
\end{figure}

Finally, the example (\ref{RGASdef}) allows us to also demonstrate conditions related to the radius of convergence of the series (\ref{ASseries}): following the discussion in Section~\ref{Sect:RadiusConvergence}, the radius of convergence\footnote{This means that beyond this radius (\ref{ASseries}) is no longer a correct representation of (\ref{RGASdef}). The latter, of course, remains valid for all values of $\widetilde{u}$.} of this series is $\frac{1}{\sqrt{|\widetilde{\omega}|}}$ and thus for any $u_c=\frac{1-\epsilon}{\sqrt{|\widetilde{\omega}|}}<\frac{1}{\sqrt{|\widetilde{\omega}|}}$ (with an infinitesimal $\epsilon>0$) we can write the coefficients $\kappa_n$ in the following form
\begin{align}
\kappa_n=\left\{\begin{array}{lcl}0 & \text{if} & n\in\mathbb{N}_{\text{odd}}\,, \\ (-(1-\epsilon)\,\text{sign}(\widetilde{\omega}))^{n/2} & \text{if} & n\in\mathbb{N}_{\text{even}}\,.\end{array}\right.
\end{align}
Comparing with the asymptotic constraints in Section~\ref{EC2}, they are in fact satisfied for $\widetilde{\omega}>0$. Notice that in this case we also have for the leading term for $\mathfrak{c}_2$ at $u_c$ in (\ref{Relc21})
\begin{align}
\mathfrak{c}_2=\frac{(6-\epsilon)(1-\epsilon)^4\chi}{4\pi (2-\epsilon)^3|\widetilde{\omega}|^{3/2}}+\mathcal{O}(u_c^4)\,,
\end{align}
while the subleading term is plotted for $\widetilde{\omega}=\frac{118}{15\pi}$ in Figure~\ref{Fig:ConditionRelRadiusConvergence}.

\section{Conclusions}\label{Sect:Conclusions}

In this paper we have investigated the implications of various positivity conditions on spherically symmetric and static geometries that are deformations of the Schwarzschild space-time. These deformations are characterised by two functions $\phi,\psi$ that enter the metric (\ref{metrica adimensionale}) through the functions $f,h$ as in eq.~(\ref{MetricFctGen}). The latter have, for example, been studied in the context of quantum models of black holes \cite{dymnikova1992vacuum,Bonanno:2000ep,Bjerrum-Bohr:2002fji,Gonzalez:2015upa,Nicolini:2019irw,Ruiz:2021qfp,Binetti:2022xdi}.  However, the results of our work apply equally to a wider class of generalisations of the Schwarzschild metric. In fact, we only require that in an asymptotic region (\emph{i.e.} sufficiently far away from the rotational center), the geometry can be described by a metric, which, via the Einstein equation (\ref{EQT}), allows to define an energy momentum tensor ${T^\mu}_\nu$, satisfying 'reasonable' physicality conditions. Concretely, by the latter we mean the above mentioned positivity conditions for the energy density or energy flux measured by physical observers \cite{Hawking:1973uf,Visser:1996iw,Ford:1994bj,Martin-Moruno:2013wfa}, reviewed in detail in Section~\ref{Sect:ConditionsT}.

We have first analysed the impact of these conditions on the metric deformations, keeping $\phi,\psi$ generic, focusing mostly on point-wise energy conditions. While most of the latter are rather difficult to reduce to a simple compact form, a common condition is $h(z)\leq f(z)$ (\emph{i.e.} $\psi(z)\geq \phi(z)$) for all values of $z$ for which ${T^\mu}_\nu$ can be defined in a meaningful manner. The analysis is simplified for the particular case $f=h$ (\emph{i.e.} $\psi=\phi$).  Indeed, in this case the conditions imply (see eq.~(\ref{NECsimple})) that $\phi=\psi\leq 1$ and $\phi'(z)\geq 0$, \emph{i.e.} the only remaining deformation function $\phi$ tends to $1$ (representing the undeformed Schwarzschild case) in a monotonic fashion, as shown schematically in Figure~\ref{Fig:SchemFuncPhi}. In the case where the deformation represents a black hole and ${T^\mu}_\nu$ (that abides to the positivity conditions) can be defined all the way to the (outer) event horizon ($\za=\zh$), this condition implies that the position $\zh$ of the latter is smaller than the one of the classical Schwarzschild black hole ($\zh\leq 2\chi$). While some conditions (such as the NEC and WEC) make no further restrictions on the growth of the function $\phi$, others (like the SEC) either impose that $\phi$ has no inflection point (an example of which is schematically shown in Figure~\ref{Fig:SchemFuncPhiSEC}) or (like the FEC or the DEC) impose that the function $\phi$ can not tend to zero too quickly as $z\to \infty$.

The (point-wise) energy conditions can be expressed in a more concrete form by assuming that the deformation functions $\phi,\psi$ are series expansions in inverse powers of the physical distance $d$ (see the definition in (\ref{DefDistanceAsymp})), as in eq.~(\ref{Functionsfh}). The precise form of the distance is conceptually of little importance: we have for example explained in Appendix~\ref{App:SchemeDependence} that shifting $d$ by a constant corresponds to a linear transformation among the expansion coefficients. More complicated changes in the distance shall be discussed elsewhere \cite{Scheme}. Analysing the leading contributions for the asymptotic limit, the positivity condition are realised only for specific ranges of the (leading) expansion coefficients $(\omega_n,\gamma_n)$, which are summarised in eq.~(\ref{WNSDECsub}) and (\ref{FECsub}) and graphically represented in Figures~\ref{Fig:LeadingOrder} and \ref{Fig:LeadingOrderFEC}. Going beyond the leading asymptotic expansion is technically challenging, however, still leads to non-trivial conditions for the expansion coefficients~(\ref{Functionsfh}). These, however, are more involved and also depend on the distance up to which the expansion is valid. Furthermore, we have distinguished regions in the parameter space of $(\omega_n,\gamma_n)$ that correspond to different minimal realisations of the energy momentum tensor in terms of physical fields: indeed, following \cite{Boonserm:2015aqa}, in different regions of the parameter space ${T^\mu}_\nu$ can be realised through either an electric vector field or a massless scalar along with a perfect fluid.

Finally, to complement our general considerations, we have also discussed  two concrete models that have been proposed as metrics for quantum black holes, namely a model proposed by Bonanno-Reuter \cite{Bonanno:2000ep} and the Dymnikova space-time \cite{dymnikova1992vacuum}. In both cases, we have found that certain energy conditions are not realised (in agreement with our previous general considerations). However, in both cases, the violation  tends to zero when going away from the black hole. Such a behaviour is not surprising in a context where quantum effects play a role: indeed, the energy conditions discussed in the bulk of this paper are inspired from classical systems and there are known quantum system in which they are violated  \cite{Epstein:1965zza,Fewster:2002ne,Visser:1996iv}. The fact that in both cases the violation of these conditions decreases (either as a power-law in the case of the Bonanno-Reuter model or exponentially in the case of the Dymnikova space-time) at infinity is compatible with the geometry approaching a classically flat space-time. In a larger context, these examples showcase that the conditions and results obtained in this paper are not to be understood as immovable 'constraints' for space-time metric but rather as guiding principles for analysing deformations of (classical) spherically symmetric metrics. We also envision these conditions to be important for estimating the size of quantum effects in astrophysical observations in the future, by providing orders of magnitude for the violation of classical energy conditions. For a more detailed discussion of quantum effects, however, in particular in the context of black holes, it might be necessary to introduce more appropriate 'quantum' conditions such as \cite{Ford:1994bj,Ford:1997fa,Bousso:2015mna,Bousso:2015wca,Kontou:2020bta}.

For future work it will be interesting to extend the analysis to more general space-times: on the one hand this includes space-times that are not spherically symmetric and/or static (\emph{e.g.} generalised rotating or charged black holes). On the other hand, this also pertains to space-times that are not asymptotically flat. In this context, we foresee interesting applications of our approach to AdS spaces. Finally, to extend our analysis, in particular in the context of black holes, to regions that are closer to the rotational center (and thus also the black hole horizon) requires probably an alternative approach. Indeed, the energy conditions we employed here are rooted in the classical behaviour of an energy momentum tensor. To define the latter, we have to assume that the metric satisfies the Einstein-equation (\ref{EQT}) at least approximately. In close proximity to the black hole, however, the latter needs to be replaced by a description rooted in quantum gravity, which is currently  unavailable. We remark, however, that even from the asymptotic approach adopted in this work, we have managed to extract non-trivial conditions for the metric functions. It would be interesting to discuss, how these could be extrapolated to gain information about regions in which quantum effects start playing a role.

\section*{Acknowledgements}
{We are highly indebted to Manuel Del Piano for collaboration at early stages of this work and related topics, as well as numerous enlightening and stimulating discussions. Similarly, we would like to thank Emanuele Binetti, Aaron Held, Giuseppe Meluccio, Alessia Platania and Mat\'ias Torres Sandoval for many useful discussions and exchanges. SH would like to thank the Munich Institute for Astro-, Particle and BioPhysics (MIAPbP) which is funded by the Deutsche Forschungsgemeinschaft (DFG, German Research Foundation) under Germany´s Excellence Strategy – EXC-2094 – 390783311 for kind support and hospitality during part of this research. FS gladly thanks the CERN theoretical physics department for the support and hospitality during the initial stages of the project.}


\appendix 
\section{General Properties of Spherical Space-Times}\label{App:SpaceTime}
\subsection{Geometry of Static, Spherically Space-Times}\label{App:Geometry}
For convenience, we record a number of geometric properties of the static, spherically symmetric metric~(\ref{metrica adimensionale}) in this appendix. For concreteness, we limit ourselves to the case $z>z_H$ (such that $f,h>0$). The Einstein tensor associated with (\ref{metrica adimensionale}) takes the form
\begin{align}
&T_{\mu\nu}=\frac{1}{8\pi}\,G_{\mu\nu}=\left(\begin{array}{cccc} h\,\epsilon & 0 & 0 & 0  \\ 0 & \frac{p_{||}}{f}& 0 & 0 \\ 0 & 0 & p_\perp\,z^2 & 0 \\ 0 & 0 & 0 & p_\perp\,z^2\,\sin^2\theta \end{array}\right)\,,&&\text{with} &&\begin{array}{l}\epsilon=  \frac{1-f-zf'}{8 \pi   z^2} \\ p_{||}=\frac{f-1+zf\frac{h'}{h}}{8 \pi   z^2} \,,\\ p_\perp =  \frac{z}{2}p'_{\parallel}+\frac{zh'}{4h}(\epsilon+p_\parallel)+p_\parallel\,.\end{array}\label{FormT}
\end{align}
The mixed tensor ${T^{\mu}}_\nu=T^{\mu\lambda}g_{\nu\lambda}$ has eigenvectors
\begin{align}
&u^\mu=\left(\begin{array}{c}-\frac{1}{\sqrt{h}} \\ 0 \\ 0 \\ 0\end{array}\right)\,,&&w^\mu=\left(\begin{array}{c} 0 \\ \sqrt{f} \\ 0 \\ 0\end{array}\right)\,,&&v_1^\mu=\left(\begin{array}{c}0 \\ 0 \\ \frac{1}{z} \\ 0\end{array}\right)\,,&&v_2^\mu=\left(\begin{array}{c}0 \\ 0 \\ 0 \\ \frac{1}{z\,\sin\theta} \end{array}\right)\,.\label{Eigenvectors}
\end{align}
which correspond to the Lorentz-invariant \cite{Martin-Moruno:2013wfa} eigenvalues $(-\epsilon,p_{||},p_\perp,p_\perp)$. These vectors are orthogonal to each other and satisfy
\begin{align}
&g_{\mu\nu}\,u^\mu u^\nu=-1\,,&&g_{\mu\nu}\,w^\mu w^\nu=g_{\mu\nu}\,v_1^\mu v_1^\nu=g_{\mu\nu}\,v_2^\mu v_2^\nu=1\,.
\end{align} 
The energy momentum tensor (\ref{FormT}) is therefore of type I in the classification of Ellis and Hawking (see Section~\ref{App:EllisHawking}). Furthermore, it can be expanded in its eigenvectors as shown in (\ref{emtensor}).

We next solve the geodesic equation
\begin{align}
&0=\frac{d \geo^\mu}{ds}+\Gamma^\mu_{\nu\lambda}\,\geo^\nu\,\geo^\lambda\,,&&\text{with}&&\geo^\mu=\left(\begin{array}{c}d\tau/ds \\ dz/ds \\ d\theta/ds \\ d\varphi/ds\end{array}\right)\,,\label{Geodesic}
\end{align}
where $s$ is an affine parameter and $\Gamma^\mu_{\nu\lambda}$ are the connection components. In the main part of this paper we are mostly interested in radial time-like and null geodesics and we therefore solve (\ref{Geodesic}) with the time-like or null initial conditions imposed at $(0,z_0\to \infty,\tfrac{\pi}{2},0)$:
{\allowdisplaybreaks
\begin{align}
&\bullet\, \text{null radial geodesic:}&&\geo^\mu(z)=\left(\begin{array}{c}\frac{1}{h(z)} \\ \sqrt{\tfrac{f(z)}{h(z)}} \\ 0 \\ 0\end{array}\right)\,,\label{NullGeo}\\
&\bullet\, \text{time-like geodesic:}&&\geo^\mu(z)=\left(\begin{array}{c}\frac{\sqrt{1+\beta^2}}{h(z)} \\ \sqrt{\tfrac{f(z)}{h(z)}(1-h(z)+\beta^2)} \\ 0 \\ 0\end{array}\right)\,,\label{TimeLikeGeo}
\end{align}}
where $\beta$ is the initial (radial) veclocity (with $\beta^2<1$).
\subsection{Radial Distance for Spherically Symmetric Metrics}\label{App:Distance}
In this appendix we provide a series expansion for the distance $d$ as a function of the radial coordinate $z$ in the asymptotic regime. The former is defined through the differential equation (\ref{DefDistanceAsymp}). We have verified up to order $\mathcal{O}(z^{-7})$ that the solution can be expanded in the form
\begin{align}
d(z)=z+\chi\left(\pi-1+\log\frac{2z}{\chi}\right)+\sum_{n=0}^\infty z^{-1-n}\sum_{j=0}^n\lambda_{j,n}\,(\log z)^j\,,\label{DefSeriesExpansion}
\end{align}
where $\lambda_{j,n}$ are $z$-independent constants, which depend on the coefficients $\omega_n$. The first few instances can be listed as follows
{\allowdisplaybreaks
\begin{align}
&\lambda_{0,0}=-\frac{\chi(3\chi+2\omega_1)}{2}\,,\hspace{0.2cm}\lambda_{0,1}=-\frac{\chi}{4}\left(5\chi^2+2\omega_2-\chi\omega_1(2\pi-7+\log 4)+2\chi\omega_1\log\chi\right)\,,\hspace{0.2cm}\lambda_{1,1}=\frac{\chi^2\omega_1}{2}\,,\nonumber\\
&\lambda_{0,2}=\frac{\chi}{216}\big[-72\omega_3-\chi\left(315\chi^2+180\omega_1^2+24\omega_2(13-6\pi-\log 64)\right)+8\chi\omega_1\big(104+\log 8(\log 8-13)\nonumber\\
&\hspace{2cm}+3\pi(3\pi-13+\log 64)\big)+24\chi\log \chi\left(-6\omega_2+\chi\omega_1(6\pi-13+\log 64)-3\chi\omega_1\log\chi\right)\big]\,,\nonumber\\
&\lambda_{1,2}=\frac{\chi^2}{9}\left(6\omega_2-\chi\omega_1(6\pi-13+\log 64)+6\chi\omega_1\,\log\chi\right)\,,\hspace{2cm}\lambda_{2,2}=-\frac{\chi^3\omega_1}{3}\,.
\end{align}}
We have verified up to order $\mathcal{O}(d^{-4})$ that the relation (\ref{DefSeriesExpansion}) can be inverted by the expansion
\begin{align}
z=d-\chi\left(\pi-1+\log\frac{2d}{\chi}\right)+\sum_{n=0}^\infty d^{-1-n}\sum_{j=0}^{n+1}\kappa_{j,n}(\log d)^j\,,\label{Expansionzd}
\end{align}
where $\kappa_{j,n}$ are $d$-independent constants, for which the first examples can be listed as
{\allowdisplaybreaks
\begin{align}
&\kappa_{0,0}=\frac{\chi}{2}\left(2\omega_1+\chi(1+2\pi+\log 4)-2\chi\log \chi\right)\,,\hspace{0.5cm} \kappa_{1,0}=\chi^2\,,\nonumber\\
&\kappa_{0,1}=\frac{\chi}{4}\big[2\omega_2+\chi\omega_1(2\pi-1+\log 4)+\chi^2(2\pi^2-1+2(\log2)^2-\log 4+\pi(\log16-2))\nonumber\\
&\hspace{2cm}-2\chi(\omega_1+\chi(2\pi-1+\log 4)\log\chi+2\chi^2(\log\chi)^2)\big]\,,\nonumber\\
&\kappa_{1,1}=\frac{\chi^2}{2}\left(\omega_1+\chi(2\pi-1+\log 4)-2\chi\log\chi\right)\,,\hspace{1cm}\kappa_{2,1}=\frac{\chi^3}{2}\,.
\end{align}}

\subsection{Classification of Energy Momentum Tensors}\label{App:EllisHawking}
Stress energy tensors have been classified locally in \cite{Hawking:1973uf} according to the degree to which they are diagonalisable through Lorentz transformations. Using the notation of \cite{Martin-Moruno:2013wfa}, at a given point in space-time there are the following four canonical types
\begin{itemize}
\item type I: In an orthonormal eigenbasis, the energy momentum tensor takes the form
\begin{align}
T^{\mu\nu}=\left(\begin{array}{cccc} \rho & 0 & 0 & 0 \\ 0 & p_1 & 0 & 0 \\ 0 & 0 & p_2 & 0 \\ 0 & 0 & 0 & p_3\end{array}\right)\,.
\end{align}
The mixed tensor ${T^{\mu}}_\nu$ has one time-like eigenvector.
\item type II: In an orthonormal eigenbasis, the energy momentum tensor takes the form
\begin{align}
T^{\mu\nu}=\left(\begin{array}{cccc} \kappa +\mathfrak{f}  & \mathfrak{f} & 0 & 0 \\ \mathfrak{f} & -\kappa +\mathfrak{f} & 0 & 0  \\ 0 & 0 & p_2 & 0 \\ 0 & 0 & 0 & p_3\end{array}\right)\,.
\end{align}
The mixed tensor ${T^{\mu}}_\nu$ has two null eigenvectors. According to \cite{Hawking:1973uf}, such tensors occur in the context of (simple) classical field configurations only for radiation travelling in the same (null) direction.
\item type III: In an orthonormal eigenbasis, the energy momentum tensor takes the form
\begin{align}
T^{\mu\nu}=\left(\begin{array}{cccc} \rho  & \tfrac{\mathfrak{f}}{\sqrt{2}} & \tfrac{\mathfrak{f}}{\sqrt{2}} & 0 \\ \tfrac{\mathfrak{f}}{\sqrt{2}} & -\rho +\mathfrak{f} & 0 & 0  \\ \tfrac{\mathfrak{f}}{\sqrt{2}} & 0 & -\rho-\mathfrak{f} & 0 \\ 0 & 0 & 0 & p_3\end{array}\right)\,.
\end{align}
The mixed tensor ${T^{\mu}}_\nu$ has three null eigenvectors. According to \cite{Hawking:1973uf}, such tensors are not sourced by (simple) classical fields.
\item type IV: In an orthonormal eigenbasis, the energy momentum tensor takes the form
\begin{align}
T^{\mu\nu}=\left(\begin{array}{cccc} \rho  & \mathfrak{f} & 0 & 0 \\ \mathfrak{f} & -\rho & 0 & 0  \\ 0 & 0 & p_2 & 0 \\ 0 & 0 & 0 & p_3\end{array}\right)\,.
\end{align}
The mixed tensor ${T^{\mu}}_\nu$ has neither time-like nor null eigenvectors. According to \cite{Hawking:1973uf}, such tensors are not sourced by (simple) classical fields.
\end{itemize}

\subsection{The Raychaudhuri Equation}\label{App:Raychaudhuri}
We define a \emph{congruence} as a set of curves within a region of space-time such that there is exactly one geodesic passing through each point of this region \cite{Hawking:1973uf}. If all curves in this family are either time-like or null, we shall call the congruence time-like or null respectively. The Raychaudhuri equation \cite{Raychaudhuri:1953yv} describes how congruences change when following geodesic motion. Our notation in the following follows mostly \cite{,Rastgoo:2022mks}.
\subsubsection{Time-like Congruences}
Consider first a congruence of time-like geodesics, whose (time-like) tangent vectors are denoted by the set $\{\xi^\mu(\tau)\}$, where $\tau$ is the proper time along the geodesic
\begin{align}
&g_{\mu\nu}\xi^\mu \xi^\nu=-1\,,&&\text{and} &&\xi^\mu \nabla_\mu \xi^\nu=0\,.
\end{align}
We can furthermore define the transverse metric $h_{\mu\nu}$ as well as the expansion tensor $B_{\mu\nu}$
\begin{align}
&h_{\mu\nu}=g_{\mu\nu}+\xi_\mu \xi_\nu\,,&&\text{and} &&B_{\mu\nu}=\nabla_\nu \xi_\mu\,.
\end{align}
The transverse metric satisfies ${h^\mu}_\mu=h_{\mu\nu}g^{\mu\nu}=3$, while the expansion tensor is orthogonal to $\xi^\mu$, \emph{i.e.} $B_{\mu\nu}\xi^\nu=0=B_{\mu\nu}\xi^\mu$. The expansion tensor can furthermore be decomposed into a trace part, a symmetric trace-less and an anti-symmetric tensor
\begin{align}
&B_{\mu\nu}=\frac{1}{3}\,\vartheta\,h_{\mu\nu}+\sigma_{\mu\nu}+\omega_{\mu\nu}\,,&&\text{with} &&\begin{array}{ll}\vartheta={B^\mu}_\mu=B_{\mu\nu}g^{\mu\nu}&\text{expansion scalar}\,,\\[4pt] \sigma_{\mu\nu}=\frac{1}{2}(B_{\mu\nu}+B_{\nu\mu})-\frac{1}{3}\,\vartheta\,h_{\mu\nu} & \text{shear tensor}\,,\\[4pt] \omega_{\mu\nu}=\frac{1}{2}(B_{\mu\nu}-B_{\nu\mu}) & \text{vorticity tensor}\,.\end{array}
\end{align}
The expansion scalar encodes the change in the area (of the cross-section) of the congruence, the shear-tensor measures the deformation of the congruence (relative to a sphere) while the vorticity tensor is a measure for the rotation of the congruence as functions of the proper time $\tau$. Indeed, these changes are encoded in Raychaudhuri equation \cite{Raychaudhuri:1953yv}
\begin{align}
\xi^\mu\nabla_\mu \vartheta=\frac{d\vartheta}{d\tau}=-\frac{1}{3}\,\vartheta^2-\sigma_{\mu\nu}\sigma^{\mu\nu}+\omega_{\mu\nu}\omega^{\mu\nu}-R_{\mu\nu}\xi^\mu \xi^\nu\,.\label{Raychaudhuritime}
\end{align}
The shear- and vorticity tensors are spatial tensors (such that $\sigma_{\mu\nu}\sigma^{\mu\nu}>0$ and $\omega_{\mu\nu}\omega^{\mu\nu}>0$). A geometric interpretation of the SEC (\emph{i.e.} $R_{\mu\nu}\xi^\mu \xi^\nu\geq 0$) is to say that only the vorticity parameter $\omega_{\mu\nu}\omega^{\mu\nu}$ can contribute to the divergence of the congruence, while all remaining terms in (\ref{Raychaudhuritime}) contribute to its convergence.
\subsubsection{Null Congruences}
The above discussion can be generalised to congruences of null geodesics: let $\{\zeta^\mu(\lambda)\}$ be their (null) tangent vectors, parametrised by $\lambda$ and let $\ell_\mu$ be an (auxiliary) null vector field
\begin{align}
&g_{\mu\nu}\zeta^\mu \zeta^\nu=0=g_{\mu\nu}\ell^\mu \ell^\nu\,,&&\text{and} &&\zeta_\mu\ell^\mu =-1\,.
\end{align}
We can then define the transverse metric $h_{\mu\nu}$ and the expansion tensor $B_{\mu\nu}$ as
\begin{align}
&h_{\mu\nu}=g_{\mu\nu}+\zeta_\mu \ell_\nu+\zeta_{\nu}\ell_\mu\,,&&\text{and} &&B_{\mu\nu}=\nabla_\nu \zeta_\mu\,.
\end{align}
The transverse metric satisfies ${h^\mu}_\mu=h_{\mu\nu}g^{\mu\nu}=2$ and can be used to define the purely transverse part of $B_{\mu\nu}$ which we can decompose into a trace part, a symmetric trace-less and an anti-symmetric tensor
\begin{align}
&\tilde{B}_{\mu\nu}:=B_{\alpha\beta}{h^\alpha}_\mu {h^\beta}_\nu=\frac{1}{2}\,\tilde{\vartheta}\,h_{\mu\nu}+\tilde{\sigma}_{\mu\nu}+\tilde{\omega}_{\mu\nu}\,,
\end{align}
where as before we introduce the expansion scalar, shear tensor and vorticity tensor
\begin{align}
&\tilde{\vartheta}={\tilde{B}^\mu}_\mu\,,&&\tilde{\sigma}_{\mu\nu}=\frac{1}{2}(\tilde{B}_{\mu\nu}+\tilde{B}_{\nu\mu})-\frac{1}{2}\,\tilde{\vartheta}\,h_{\mu\nu} \,,&&\tilde{\omega}_{\mu\nu}=\frac{1}{2}(\tilde{B}_{\mu\nu}-\tilde{B}_{\nu\mu}) \,.
\end{align}
The Raychaudhuri equation \cite{Raychaudhuri:1953yv} describes the evolution of $\tilde{\vartheta}$ along a geodesic
\begin{align}
\zeta^\mu\nabla_\mu \tilde{\vartheta}=\frac{d\tilde{\vartheta}}{d\lambda}=-\frac{1}{2}\,\tilde{\vartheta}^2-\tilde{\sigma}_{\mu\nu}\tilde{\sigma}^{\mu\nu}+\tilde{\omega}_{\mu\nu}\tilde{\omega}^{\mu\nu}-R_{\mu\nu}\zeta^\mu \zeta^\nu\,.\label{Raychaudhurinull}
\end{align}
Both $\tilde{\theta}$ as well as (\ref{Raychaudhurinull}) are independent of the choice of $\ell_\mu$.

\section{ On the physical distance choice }\label{App:SchemeDependence}
In eq.~(\ref{DefDistanceAsymp}) we have defined the proper distance to the rotational center of the metric in a fashion, which is slightly different from previous work (see \emph{e.g.} \cite{Bonanno:2000ep,Binetti:2022xdi}). Concretely, the definition~(\ref{DefDistanceAsymp}) in terms of a first-order differential equation leaves the ambiguity of an additive constant (which is fixed by the second equation in (\ref{DefDistanceAsymp})). In this appendix, we shall argue that shifting the proper distance by a (position independent) constant, has no conceptual implications on the main conclusions of this paper, for example the conditions (\ref{LeadingDWNSEC}) and (\ref{FECleadingCond}) (or their subleading counterparts (\ref{WNSDECsub}) and (\ref{FECsub})).

Concretely, we consider the series expansions (\ref{Functionsfh}) for $u(z)=\frac{1}{d(z)}\in[0,u_c)$, with $u_c>0$ the radius of convergence. We then implement a shift of $d$ by a constant, which we shall call $\beta\in\mathbb{R}$, \emph{i.e.} $d(z)\to d(z)+\beta$. In the variable $u$, we therefore have
\begin{align}
u\longrightarrow \sigma(u)=\frac{u}{1+\beta\,u}=\sum_{k=1}^\infty c_k\,u^k\,,&&\text{with} &&c_k=(-\beta)^{k-1}\,.
\end{align} 
This series expansion has radius of convergence $1/|\beta|$, such that the linear change of variables $\omega'_n\to \omega_n$ is only valid for $|\beta|\leq d_c$, \emph{i.e.} we can not shift the distance function by constants that are larger than the minimal distance to the black hole for which we trust the series (\ref{Functionsfh}) to still be a valid description of the metric function. In this case, we can write for the metric function~$f$\footnote{A similar discussion also applies to the function $h$.}
\begin{align}
f(z)=1-\frac{2\chi}{z}\sum_{n=0}^\infty \omega'_n\,\sigma^n\stackrel{!}{=}1-\frac{2\chi}{z}\sum_{n=0}^\infty \omega_n\,u^n
\end{align}
which implies the following relation among the coefficients $\omega_n$ and $\omega'_n$.
\begin{align}
\left(\begin{array}{c}\omega_0 \\ \omega_1 \\ \omega_2 \\ \omega_3 \\ \omega_4 \\ \omega_5\\ \vdots\end{array}\right)=\left(\begin{array}{ccccccc}1 & 0 & 0 & 0 & 0& 0 & \cdots \\ 0 & 1 & 0 & 0 & 0 & 0 & \cdots \\ 0 & -\beta & 1 & 0 & 0 & 0 & \cdots \\ 0 & \beta^2 & -2\beta & 1 & 0 & 0 & \cdots \\ 0 & -\beta^3 & 3\beta^2 & -3\beta & 1 &  0 & \cdots \\ 0 & \beta^4 & -4\beta & 6\beta^2 & -4\beta & 1 & \cdots\\ \vdots & \vdots & \vdots & \vdots &\vdots & \vdots & \ddots\end{array}\right)\cdot\left(\begin{array}{c}\omega'_0 \\ \omega'_1 \\ \omega'_2 \\ \omega'_3 \\ \omega'_4 \\ \omega'_5\\ \vdots\end{array}\right)\,,\label{GenLinTransform2}
\end{align}
which can be written explicitly
\begin{align}
&\omega_0=\omega'_0\,,&&\text{and} &&\omega_n=\sum_{m=0}^{n-1}\left(\begin{array}{c} n-1 \\ m\end{array}\right)\,(-\beta)^m\,\omega'_{n-m}\hspace{0.5cm}\forall n\geq 1\,.\label{linRelOms}
\end{align}
Notice that the shift of $d$ by a (finite) constant implies to replace the expansion coefficients $\omega_n$ in the metric functions (\ref{Functionsfh}) by \emph{finite} linear combinations of these coefficients. Notice also that the inverse of the replacement (\ref{linRelOms}) is obtained by replacing $\beta\to -\beta$.
 
Finally, we remark that transformations of the distance function that are more general than a shift by a simple constant, can be discussed in a similar fashion. Such \emph{scheme changes} shall be discussed elsewhere \cite{Scheme}.

\printbibliography

@article{Binetti:2022xdi,
    author = "Binetti, Emanuele and Del Piano, Manuel and Hohenegger, Stefan and Pezzella, Franco and Sannino, Francesco",
    title = "{The Effective Theory of Quantum Black Holes}",
    eprint = "2203.13515",
    archivePrefix = "arXiv",
    primaryClass = "gr-qc",
    reportNumber = "CERN-TH-2022-048",
    month = "3",
    year = "2022"
}

@article{Boonserm:2015aqa,
    author = "Boonserm, Petarpa and Ngampitipan, Tritos and Visser, Matt",
    title = "{Mimicking static anisotropic fluid spheres in general relativity}",
    eprint = "1501.07044",
    archivePrefix = "arXiv",
    primaryClass = "gr-qc",
    doi = "10.1142/S021827181650019X",
    journal = "Int. J. Mod. Phys. D",
    volume = "25",
    number = "02",
    pages = "1650019",
    year = "2015"
}

@article{Kontou:2020bta,
    author = "Kontou, Eleni-Alexandra and Sanders, Ko",
    title = "{Energy conditions in general relativity and quantum field theory}",
    eprint = "2003.01815",
    archivePrefix = "arXiv",
    primaryClass = "gr-qc",
    doi = "10.1088/1361-6382/ab8fcf",
    journal = "Class. Quant. Grav.",
    volume = "37",
    number = "19",
    pages = "193001",
    year = "2020"
}

@article{Bonanno:2000ep,
    author = "Bonanno, Alfio and Reuter, Martin",
    title = "{Renormalization group improved black hole space-times}",
    eprint = "hep-th/0002196",
    archivePrefix = "arXiv",
    reportNumber = "INFN-CT-03-00, MZ-TH-00-04",
    doi = "10.1103/PhysRevD.62.043008",
    journal = "Phys. Rev. D",
    volume = "62",
    pages = "043008",
    year = "2000"
}

@article{Cadoni:2022chn,
    author = "Cadoni, Mariano and Oi, Mauro and Sanna, Andrea Pierfrancesco",
    title = "{Effective models of nonsingular quantum black holes}",
    eprint = "2204.09444",
    archivePrefix = "arXiv",
    primaryClass = "gr-qc",
    reportNumber = "Phys. Rev. D 106, 024030",
    doi = "10.1103/PhysRevD.106.024030",
    journal = "Phys. Rev. D",
    volume = "106",
    number = "2",
    pages = "024030",
    year = "2022"
}

@article{Abreu:2011fr,
    author = "Abreu, Gabriel and Barcelo, Carlos and Visser, Matt",
    title = "{Entropy bounds in terms of the w parameter}",
    eprint = "1109.2710",
    archivePrefix = "arXiv",
    primaryClass = "gr-qc",
    doi = "10.1007/JHEP12(2011)092",
    journal = "JHEP",
    volume = "12",
    pages = "092",
    year = "2011"
}

@article{dymnikova1992vacuum,
  title={Vacuum nonsingular black hole},
  author={Dymnikova, Irina},
  journal={General relativity and gravitation},
  volume={24},
  number={3},
  pages={235--242},
  year={1992},
  publisher={Springer}
}

@ARTICLE{Schwarzschild,
       author = {{Schwarzschild}, Karl},
        title = "{{\"U}ber das Gravitationsfeld eines Massenpunktes nach der Einsteinschen Theorie}",
      journal = {Sitzungsberichte der K{\"o}niglich Preussischen Akademie der Wissenschaften},
         year = 1916,
        month = jan,
        pages = {189-196},
}

@article{Page:1982fm,
    author = "Page, Don N.",
    title = "{Thermal Stress Tensors in Static Einstein Spaces}",
    reportNumber = "PRINT-82-0258 (PENN-STATE)",
    doi = "10.1103/PhysRevD.25.1499",
    journal = "Phys. Rev. D",
    volume = "25",
    pages = "1499",
    year = "1982"
}

@article{DEWITT1975295,
title = {Quantum field theory in curved spacetime},
journal = {Physics Reports},
volume = {19},
number = {6},
pages = {295-357},
year = {1975},
issn = {0370-1573},
doi = {https://doi.org/10.1016/0370-1573(75)90051-4},
url = {https://www.sciencedirect.com/science/article/pii/0370157375900514},
author = {Bryce S. DeWitt}
}

@book{Birrell:1982ix,
    author = "Birrell, N. D. and Davies, P. C. W.",
    title = "{Quantum Fields in Curved Space}",
    doi = "10.1017/CBO9780511622632",
    isbn = "978-0-521-27858-4, 978-0-521-27858-4",
    publisher = "Cambridge Univ. Press",
    address = "Cambridge, UK",
    series = "Cambridge Monographs on Mathematical Physics",
    month = "2",
    year = "1984"
}

@article{Brown:1985ri,
    author = "Brown, M. R. and Ottewill, A. C.",
    title = "{EFFECTIVE ACTIONS AND CONFORMAL TRANSFORMATIONS}",
    doi = "10.1103/PhysRevD.31.2514",
    journal = "Phys. Rev. D",
    volume = "31",
    pages = "2514--2520",
    year = "1985"
}

@book{Hawking:1973uf,
    author = "Hawking, S. W. and Ellis, G. F. R.",
    title = "{The Large Scale Structure of Space-Time}",
    doi = "10.1017/CBO9780511524646",
    isbn = "978-0-521-20016-5, 978-0-521-09906-6, 978-0-511-82630-6, 978-0-521-09906-6",
    publisher = "Cambridge University Press",
    series = "Cambridge Monographs on Mathematical Physics",
    month = "2",
    year = "2011"
}

@article{Visser:1996iw,
    author = "Visser, Matt",
    title = "{Gravitational vacuum polarization. 1: Energy conditions in the Hartle-Hawking vacuum}",
    eprint = "gr-qc/9604007",
    archivePrefix = "arXiv",
    doi = "10.1103/PhysRevD.54.5103",
    journal = "Phys. Rev. D",
    volume = "54",
    pages = "5103--5115",
    year = "1996"
}

@article{Ford:1994bj,
    author = "Ford, L. H. and Roman, Thomas A.",
    title = "{Averaged energy conditions and quantum inequalities}",
    eprint = "gr-qc/9410043",
    archivePrefix = "arXiv",
    reportNumber = "TUTP-94-16",
    doi = "10.1103/PhysRevD.51.4277",
    journal = "Phys. Rev. D",
    volume = "51",
    pages = "4277--4286",
    year = "1995"
}

@article{Epstein:1965zza,
    author = "Epstein, H. and Glaser, V. and Jaffe, A.",
    title = "{Nonpositivity of energy density in Quantized field theories}",
    doi = "10.1007/BF02749799",
    journal = "Nuovo Cim.",
    volume = "36",
    pages = "1016",
    year = "1965"
}

@article{Fewster:2002ne,
    author = "Fewster, Christopher J. and Roman, Thomas A.",
    title = "{Null energy conditions in quantum field theory}",
    eprint = "gr-qc/0209036",
    archivePrefix = "arXiv",
    doi = "10.1103/PhysRevD.67.044003",
    journal = "Phys. Rev. D",
    volume = "67",
    pages = "044003",
    year = "2003",
    note = "[Erratum: Phys.Rev.D 80, 069903 (2009)]"
}

@article{Visser:1996iv,
    author = "Visser, Matt",
    title = "{Gravitational vacuum polarization. 2: Energy conditions in the Boulware vacuum}",
    eprint = "gr-qc/9604008",
    archivePrefix = "arXiv",
    doi = "10.1103/PhysRevD.54.5116",
    journal = "Phys. Rev. D",
    volume = "54",
    pages = "5116--5122",
    year = "1996"
}

@article{Barcelo:2002bv,
    author = "Barcelo, Carlos and Visser, Matt",
    title = "{Twilight for the energy conditions?}",
    eprint = "gr-qc/0205066",
    archivePrefix = "arXiv",
    doi = "10.1142/S0218271802002888",
    journal = "Int. J. Mod. Phys. D",
    volume = "11",
    pages = "1553--1560",
    year = "2002"
}

@article{Ford:1997fa,
    author = "Ford, L. H. and Pfenning, Michael J. and Roman, Thomas A.",
    title = "{Quantum inequalities and singular negative energy densities}",
    eprint = "gr-qc/9711030",
    archivePrefix = "arXiv",
    reportNumber = "TUTP-97-11",
    doi = "10.1103/PhysRevD.57.4839",
    journal = "Phys. Rev. D",
    volume = "57",
    pages = "4839--4846",
    year = "1998"
}

@article{Bousso:2015mna,
    author = "Bousso, Raphael and Fisher, Zachary and Leichenauer, Stefan and Wall, Aron C.",
    title = "{Quantum focusing conjecture}",
    eprint = "1506.02669",
    archivePrefix = "arXiv",
    primaryClass = "hep-th",
    doi = "10.1103/PhysRevD.93.064044",
    journal = "Phys. Rev. D",
    volume = "93",
    number = "6",
    pages = "064044",
    year = "2016"
}

@article{Bousso:2015wca,
    author = "Bousso, Raphael and Fisher, Zachary and Koeller, Jason and Leichenauer, Stefan and Wall, Aron C.",
    title = "{Proof of the Quantum Null Energy Condition}",
    eprint = "1509.02542",
    archivePrefix = "arXiv",
    primaryClass = "hep-th",
    doi = "10.1103/PhysRevD.93.024017",
    journal = "Phys. Rev. D",
    volume = "93",
    number = "2",
    pages = "024017",
    year = "2016"
}

@article{Martin-Moruno:2013sfa,
    author = "Martin-Moruno, Prado and Visser, Matt",
    title = "{Classical and quantum flux energy conditions for quantum vacuum states}",
    eprint = "1305.1993",
    archivePrefix = "arXiv",
    primaryClass = "gr-qc",
    doi = "10.1103/PhysRevD.88.061701",
    journal = "Phys. Rev. D",
    volume = "88",
    number = "6",
    pages = "061701",
    year = "2013"
}

@article{Martin-Moruno:2013wfa,
    author = "Martin-Moruno, Prado and Visser, Matt",
    title = "{Semiclassical energy conditions for quantum vacuum states}",
    eprint = "1306.2076",
    archivePrefix = "arXiv",
    primaryClass = "gr-qc",
    doi = "10.1007/JHEP09(2013)050",
    journal = "JHEP",
    volume = "09",
    pages = "050",
    year = "2013"
}

@article{Curiel:2014zba,
    author = "Curiel, Erik",
    title = "{A Primer on Energy Conditions}",
    eprint = "1405.0403",
    archivePrefix = "arXiv",
    primaryClass = "physics.hist-ph",
    doi = "10.1007/978-1-4939-3210-8_3",
    journal = "Einstein Stud.",
    volume = "13",
    pages = "43--104",
    year = "2017"
}

@article{Raychaudhuri:1953yv,
    author = "Raychaudhuri, Amalkumar",
    title = "{Relativistic cosmology. 1.}",
    doi = "10.1103/PhysRev.98.1123",
    journal = "Phys. Rev.",
    volume = "98",
    pages = "1123--1126",
    year = "1955"
}

@article{Rastgoo:2022mks,
    author = "Rastgoo, Saeed and Das, Saurya",
    title = "{Probing the Interior of the Schwarzschild Black Hole Using Congruences: LQG vs. GUP}",
    eprint = "2205.03799",
    archivePrefix = "arXiv",
    primaryClass = "gr-qc",
    doi = "10.3390/universe8070349",
    journal = "Universe",
    volume = "8",
    number = "7",
    pages = "349",
    year = "2022"
}

@article{Ford:1978qya,
    author = "Ford, L. H.",
    title = "{Quantum Coherence Effects and the Second Law of Thermodynamics}",
    doi = "10.1098/rspa.1978.0197",
    journal = "Proc. Roy. Soc. Lond. A",
    volume = "364",
    pages = "227--236",
    year = "1978"
}

@article{Ford:1990id,
    author = "Ford, L. H.",
    title = "{Constraints on negative energy fluxes}",
    reportNumber = "TUTP-90-2",
    doi = "10.1103/PhysRevD.43.3972",
    journal = "Phys. Rev. D",
    volume = "43",
    pages = "3972--3978",
    year = "1991"
}

@article{Fewster:2006ti,
    author = "Fewster, Christopher J. and Osterbrink, Lutz W.",
    title = "{Averaged energy inequalities for the non-minimally coupled classical scalar field}",
    eprint = "gr-qc/0606009",
    archivePrefix = "arXiv",
    doi = "10.1103/PhysRevD.74.044021",
    journal = "Phys. Rev. D",
    volume = "74",
    pages = "044021",
    year = "2006"
}

@article{Fewster:2007ec,
    author = "Fewster, Christopher J. and Osterbrink, Lutz W.",
    title = "{Quantum Energy Inequalities for the Non-Minimally Coupled Scalar Field}",
    eprint = "0708.2450",
    archivePrefix = "arXiv",
    primaryClass = "gr-qc",
    doi = "10.1088/1751-8113/41/2/025402",
    journal = "J. Phys. A",
    volume = "41",
    pages = "025402",
    year = "2008"
}

@inproceedings{Roman:2004xm,
    author = "Roman, Thomas A.",
    title = "{Some thoughts on energy conditions and wormholes}",
    booktitle = "{10th Marcel Grossmann Meeting on Recent Developments in Theoretical and Experimental General Relativity, Gravitation and Relativistic Field Theories (MG X MMIII)}",
    eprint = "gr-qc/0409090",
    archivePrefix = "arXiv",
    doi = "10.1142/9789812704030_0236",
    pages = "1909--1922",
    month = "9",
    year = "2004"
}

@inproceedings{Fewster:2003vg,
    author = "Fewster, Christopher J.",
    title = "{Energy inequalities in quantum field theory}",
    booktitle = "{14th International Congress on Mathematical Physics}",
    pages = "559--568",
    month = "7",
    year = "2003"
}

@article{CosenzaHerrera,
author = {Cosenza,M.  and Herrera,L.  and Esculpi,M.  and Witten,L. },
title = {Some models of anisotropic spheres in general relativity},
journal = {Journal of Mathematical Physics},
volume = {22},
number = {1},
pages = {118-125},
year = {1981},
doi = {10.1063/1.524742},

URL = { 
        https://doi.org/10.1063/1.524742
    
},
eprint = { 
        https://doi.org/10.1063/1.524742
    
}

}

@book{Carroll:2004st,
    author = "Carroll, Sean M.",
    title = "{Spacetime and Geometry}",
    isbn = "978-0-8053-8732-2, 978-1-108-48839-6, 978-1-108-77555-7",
    publisher = "Cambridge University Press",
    month = "7",
    year = "2019"
}

@article{Donoghue:1994dn,
    author = "Donoghue, John F.",
    title = "{General relativity as an effective field theory: The leading quantum corrections}",
    eprint = "gr-qc/9405057",
    archivePrefix = "arXiv",
    reportNumber = "UMHEP-408",
    doi = "10.1103/PhysRevD.50.3874",
    journal = "Phys. Rev. D",
    volume = "50",
    pages = "3874--3888",
    year = "1994"
}

@inproceedings{Donoghue:1995cz,
    author = "Donoghue, John F.",
    title = "{Introduction to the effective field theory description of gravity}",
    booktitle = "{Advanced School on Effective Theories}",
    eprint = "gr-qc/9512024",
    archivePrefix = "arXiv",
    reportNumber = "UMHEP-424",
    month = "6",
    year = "1995"
}

@article{Gonzalez:2015upa,
    author = "Gonz\'alez, Cristopher and Koch, Benjamin",
    title = {{Improved Reissner\textendash{}Nordstr\"om\textendash{}(A)dS black hole in asymptotic safety}},
    eprint = "1508.01502",
    archivePrefix = "arXiv",
    primaryClass = "hep-th",
    doi = "10.1142/S0217751X16501414",
    journal = "Int. J. Mod. Phys. A",
    volume = "31",
    number = "26",
    pages = "1650141",
    year = "2016"
}

@article{Ruiz:2021qfp,
    author = "Ruiz, O. and Tuiran, E.",
    title = {{Non-Perturbative Quantum Correction to the Reissner-Nordstr\"om spacetime with Running Newton's Constant}},
    eprint = "2112.12519",
    archivePrefix = "arXiv",
    primaryClass = "gr-qc",
    month = "12",
    year = "2021"
}

@article{Donoghue:2022chi,
    author = "Donoghue, John F.",
    title = "{Non-local partner to the cosmological constant}",
    eprint = "2201.12217",
    archivePrefix = "arXiv",
    primaryClass = "hep-th",
    reportNumber = "ACFI-T22-01",
    month = "1",
    year = "2022"
}

@article{Bjerrum-Bohr:2002gqz,
    author = "Bjerrum-Bohr, N. E. J and Donoghue, John F. and Holstein, Barry R.",
    title = "{Quantum gravitational corrections to the nonrelativistic scattering potential of two masses}",
    eprint = "hep-th/0211072",
    archivePrefix = "arXiv",
    doi = "10.1103/PhysRevD.71.069903",
    journal = "Phys. Rev. D",
    volume = "67",
    pages = "084033",
    year = "2003",
    note = "[Erratum: Phys.Rev.D 71, 069903 (2005)]"
}

@article{Nicolini:2019irw,
    author = "Nicolini, Piero and Spallucci, Euro and Wondrak, Michael F.",
    title = "{Quantum Corrected Black Holes from String T-Duality}",
    eprint = "1902.11242",
    archivePrefix = "arXiv",
    primaryClass = "gr-qc",
    doi = "10.1016/j.physletb.2019.134888",
    journal = "Phys. Lett. B",
    volume = "797",
    pages = "134888",
    year = "2019"
}

@article{Bjerrum-Bohr:2002fji,
    author = "Bjerrum-Bohr, Niels Emil Jannik and Donoghue, John F. and Holstein, Barry R.",
    title = "{Quantum corrections to the Schwarzschild and Kerr metrics}",
    eprint = "hep-th/0211071",
    archivePrefix = "arXiv",
    doi = "10.1103/PhysRevD.68.084005",
    journal = "Phys. Rev. D",
    volume = "68",
    pages = "084005",
    year = "2003",
    note = "[Erratum: Phys.Rev.D 71, 069904 (2005)]"
}

@article{ChenGoldenfeldOono1995,
   title={Renormalization group and singular perturbations: Multiple scales, boundary layers, and reductive perturbation theory},
   volume={54},
   ISSN={1095-3787},
   url={http://dx.doi.org/10.1103/PhysRevE.54.376},
   DOI={10.1103/physreve.54.376},
   number={1},
   journal={Physical Review E},
   publisher={American Physical Society (APS)},
   author={Chen, Lin-Yuan and Goldenfeld, Nigel and Oono, Y.},
   year={1996},
   month={7},
   pages={376–394}
}

@article{PhysRevE.49.4502,
  title = {Renormalization group theory and variational calculations for propagating fronts},
  author = {Chen, Lin-Yuan and Goldenfeld, Nigel and Oono, Y.},
  journal = {Phys. Rev. E},
  volume = {49},
  issue = {5},
  pages = {4502--4511},
  numpages = {0},
  year = {1994},
  month = {5},
  publisher = {American Physical Society},
  doi = {10.1103/PhysRevE.49.4502},
  url = {https://link.aps.org/doi/10.1103/PhysRevE.49.4502}
  }

@article{Barenblatt,
author = {Barenblatt, G.I.},
year = {1976},
pages = {643-664},
title = {Self-similarity: Similarity and intermediate asymptotic form},
volume = {19},
journal = {Radiophysics and Quantum Electronics},
doi = {10.1007/BF01043552}
}

@article{Platania:2019kyx,
    author = "Platania, Alessia",
    title = "{Dynamical renormalization of black-hole spacetimes}",
    eprint = "1903.10411",
    archivePrefix = "arXiv",
    primaryClass = "gr-qc",
    doi = "10.1140/epjc/s10052-019-6990-2",
    journal = "Eur. Phys. J. C",
    volume = "79",
    number = "6",
    pages = "470",
    year = "2019"
}

@article{Held:2021vwd,
    author = "Held, Aaron",
    title = "{Invariant Renormalization-Group improvement}",
    eprint = "2105.11458",
    archivePrefix = "arXiv",
    primaryClass = "gr-qc",
    reportNumber = "Imperial/TP/2021/AH/04",
    month = "5",
    year = "2021"
}

@article{Penrose:1964wq,
    author = "Penrose, Roger",
    title = "{Gravitational collapse and space-time singularities}",
    doi = "10.1103/PhysRevLett.14.57",
    journal = "Phys. Rev. Lett.",
    volume = "14",
    pages = "57--59",
    year = "1965"
}

@article{osti_4155937,
    author = "Gannon, D",
    title = "{Singularities in nonsimply connected space--times}",
    doi = "10.1063/1.522498",
    journal = "J. Math. Phys. (N.Y.)",
    volume = "16",
    pages = "2364-2367",
    year = "1975"
}

@article{Hawking:1970zqf,
    author = "Hawking, S. W. and Penrose, R.",
    title = "{The Singularities of gravitational collapse and cosmology}",
    doi = "10.1098/rspa.1970.0021",
    journal = "Proc. Roy. Soc. Lond. A",
    volume = "314",
    pages = "529--548",
    year = "1970"
}

@article{Penrose:1969pc,
    author = "Penrose, R.",
    title = "{Gravitational collapse: The role of general relativity}",
    doi = "10.1023/A:1016578408204",
    journal = "Riv. Nuovo Cim.",
    volume = "1",
    pages = "252--276",
    year = "1969"
}

@book{Wald:1984rg,
    author = "Wald, Robert M.",
    title = "{General Relativity}",
    doi = "10.7208/chicago/9780226870373.001.0001",
    publisher = "Chicago Univ. Pr.",
    address = "Chicago, USA",
    year = "1984"
}

@incollection{Trautman1965-TRAFAC,
	pages = {1--1},
	booktitle = {Lectures on General Relativity},
	title = {Foundations and Current Problems of General Relativity (Notes by Graham Dixon, Petros Florides and Gerald Lemmer)},
	publisher = {Englewood Cliffs, N.J., Prentice-Hall},
	author = {Andrzej Trautman},
	editor = {A. Trautman},
	year = {1965}
}

@article{ATrautman_1966,
doi = {10.1070/PU1966v009n03ABEH002883},
url = {https://dx.doi.org/10.1070/PU1966v009n03ABEH002883},
year = {1966},
month = {mar},
publisher = {},
volume = {9},
number = {3},
pages = {319},
author = {A Trautman},
title = {THE GENERAL THEORY OF RELATIVITY},
journal = {Soviet Physics Uspekhi}
}

@article{PhysRevD.86.083515,
  title = {Energy conditions bounds on $f(T)$ gravity},
  author = {Liu, Di and Rebou\ifmmode \mbox{\c{c}}\else \c{c}\fi{}as, M. J.},
  journal = {Phys. Rev. D},
  volume = {86},
  issue = {8},
  pages = {083515},
  numpages = {8},
  year = {2012},
  month = {Oct},
  publisher = {American Physical Society},
  doi = {10.1103/PhysRevD.86.083515},
  url = {https://link.aps.org/doi/10.1103/PhysRevD.86.083515}
}

@article{Hayward_2006,
	doi = {10.1103/physrevlett.96.031103},
  
	url = {https://doi.org/10.1103%2Fphysrevlett.96.031103},
  
	year = 2006,
	month = {jan},
  
	publisher = {American Physical Society ({APS})},
  
	volume = {96},
  
	number = {3},
  
	author = {Sean A. Hayward},
  
	title = {Formation and Evaporation of Nonsingular Black Holes},
  
	journal = {Physical Review Letters}
}

@article{Eichhorn:2022bgu,
    author = "Eichhorn, Astrid and Held, Aaron",
    title = "{Black holes in asymptotically safe gravity and beyond}",
    eprint = "2212.09495",
    archivePrefix = "arXiv",
    primaryClass = "gr-qc",
    month = "12",
    year = "2022"
}

@article{bardeen1968proceedings,
author={Bardeen, James M},
title= {Non-singular general relativistic gravitational collapse},
  journal={Proceedings of the International Conference GR5},
  year={1968},
  publisher={Tbilisi University Press Tbilisi}
}

@article{Knorr:2022kqp,
    author = "Knorr, Benjamin and Platania, Alessia",
    title = "{Sifting quantum black holes through the principle of least action}",
    eprint = "2202.01216",
    archivePrefix = "arXiv",
    primaryClass = "hep-th",
    doi = "10.1103/PhysRevD.106.L021901",
    journal = "Phys. Rev. D",
    volume = "106",
    number = "2",
    pages = "L021901",
    year = "2022"
}

@misc{Scheme,
 author = "Del Piano, Manuel and Hohenegger, Stefan and Sannino, Francesco",
    Title = {work in progress}
}
\end{document}